\documentclass[12pt]{article}

\usepackage{scicite}

\usepackage{times}

\usepackage{amssymb}
\usepackage{color,graphicx}
\usepackage{amsmath}
\usepackage{amsbsy}
\usepackage{amsthm}
\usepackage{bbm}
\usepackage{bm}
\usepackage{epsfig}
\usepackage{lscape}
\usepackage{float}
\usepackage{subfigure}
\usepackage{hyperref}
\usepackage{ulem}

\newenvironment{sciabstract}{%
\begin{quote} \bf}
{\end{quote}}

\newcommand{\rev}[1]{{\color{black} #1}}

\title{Ultrafast Meissner-Levitated Microrotor}

\author
{Joel K Jose,$^{1}$ Andrea Marchese,$^{2,3}$ Marion Cromb,$^{4}$\\
Hendrik Ulbricht,$^{4}$ Andrejs Cebers,$^{5}$ Ping Koy Lam,$^{1}$\\
Tao Wang,$^{1,\dagger}$ Andrea Vinante,$^{2,3,\ast}$ \\
\\
\normalsize{$^{1}$Quantum Innovation Centre (Q.InC), Agency for Science Technology and Research} \\
\normalsize{(A*STAR), 2 Fusionopolis Way, Innovis \#08-03, Singapore 138634, Republic of Singapore,}\\
\normalsize{$^{2}$Istituto di Fotonica e Nanotecnologie IFN-CNR, 38123 Povo, Trento, Italy}\\
\normalsize{$^{3}$Fondazione Bruno Kessler (FBK), 38123 Povo, Trento, Italy}\\
\normalsize{$^{4}$School of Physics and Astronomy, University of Southampton, SO17 1BJ, Southampton, UK}\\
\normalsize{$^{5}$MMML lab, Department of Physics, University of Latvia, Jelgavas-3, R\={i}ga, LV-1004, Latvia.}\\
\normalsize{$^\ast$E-mail: andrea.vinante@cnr.it}
\normalsize{$^\dagger$E-mail: tao\_wang@a-star.edu}
}

\date{}

\begin{document} 


\baselineskip24pt


\maketitle


\paragraph*{\textbf{Teaser:} Record-setting magnetically levitated microrotor at 138 Mrpm enables ultrasensitive sensing and new tests of fundamental physics.}


\begin{sciabstract}
Levitated mechanical systems offer a unique and flexible platform for the realization of ultralow noise sensors and tests of fundamental physics. Here, we report the demonstration of a levitated microrotor, consisting of a magnetic microsphere levitated by the Meissner effect at 4.2 K. The rotor reaches rotation frequencies above 2.3 MHz (138 million rpm) while exhibiting an exceptionally low damping rate 4.75 $\times$10$^{-7}$ Hz. Exploiting the linear dependence of friction on the gas density, we demonstrate high-accuracy absolute pressure sensing over a wide range at cryogenic temperature. \rev{Finally, we detect the thermal torque noise at intermediate pressure by operating the rotor in synchronized mode, resulting in a first preliminary narrowband search for axion-like dark matter.} More broadly, the unique combination of high rotational speed, low temperature, and low thermal torque noise establishes a new regime for precision measurements, with direct implications for the detection of extremely weak forces and fields. 
\end{sciabstract}

\


\section{Introduction}

Magnetically levitated microparticles have achieved exceptionally low energy dissipation with mechanical Q-factors exceeding $10^7$ \cite{hofer2023high,vinante2020ultralow} and are a very promising tool for precision measurements in particular in magnetometry \cite{jackson2016precessing} and gravimetry \cite{Timberlake2019}. For example, precessing ferromagnets can potentially achieve magnetic field sensitivity well below the standard quantum limit (SQL) \cite{jackson2016precessing,ni2025microscopic}. Following initial pioneering works \cite{wang2019dynamics,vinante2020ultralow}, levitated magnetometers have demonstrated energy resolutions well below $\hbar$ \cite{Vinante2021,ahrens2025levitated}.
Levitated micromagnets also have great potential in quantum technologies. By coupling Meissner-levitated micromagnets with NV center spins, a spin-mechanical quantum interface is realized, with potential applications in quantum processors and spin-based quantum architectures \cite{gieseler2020single, fung2024toward}.


All of these applications exploit the oscillatory (translational and librational) modes of the levitated micromagnet. However, no studies have been performed so far on spinning levitated micromagnets. In contrast, macroscopic magnetically levitated rotors are a well-developed technology enabling important applications, such as turbomolecular pumps, energy storage \cite{flywheels}, pressure sensors \cite{beams1962spinning} and gyroscopes \cite{geim2001detection,gravityprobeB}. So far, these systems were limited to the macroscopic regime, relatively low frequencies and mostly to room temperature. To our knowledge, the fastest reported rotational speed for a Meissner-levitated rotor was 20 kHz, and was achieved using high $T_c$ superconductors \cite{moon1990}. The fastest magnetically levitated rotor reported so far by any means was a steel sphere of $0.5$ mm, spun at 660 kHz by feedback-stabilized levitation combined with asynchronous driving \cite{schuck2018}. 

Here, we demonstrate stable levitation and ultrafast spinning of micromagnets with radius $\sim 30 \, \mu$m. The setup operates at cryogenic temperature and employs superconductor-assisted Meissner levitation, allowing damping rates below $10^{-6}$ Hz. Moreover, we report the fastest rotational velocity achieved by a magnetically levitated rotor, measured at 2.3 MHz (138 million rpm). 


Our rotor can be operated as a cryogenic pressure sensor by measuring the decay rate of its angular velocity, similarly to conventional Spinning-Rotor Gauges (SRGs) \cite{beams1962spinning}. SRGs operate in the molecular flow regime, and exploit the proportionality between gas pressure and rotational drag induced by the gas \cite{blakemore2020absolute}. Commercial SRGs utilize a magnetically levitated sphere but do not rely on Meissner levitation, thus requiring a complex stabilization of the magnet position by three-dimensional active feedback. SRGs function as absolute pressure gauges \cite{fremerey1985spinning} and are considered a highly stable and precise transfer standard for low pressure measurements by NIST and other national metrology institutes \cite{fedchak2015recommended}. 
Our system, that we call Meissner-levitated spinning rotor (MLSR), encompasses all the advantages of an SRG but can operate at Kelvin and potentially even lower temperatures, enabling accurate measurements of pressure inside cryostats. This is relevant for emerging quantum technologies relying on mechanical systems, where gas collisions can be a relevant source of decoherence. 

Besides practical applications, such as pressure sensing in a novel regime, our rotor features an extremely low thermal torque noise, due to the unique combination of ultralow damping and low temperature. This offers novel opportunities in precision measurements relevant to fundamental physics, such as the search for dark matter, high-frequency gravitational waves, and the test of the limits of quantum mechanics. \rev{As a concrete example, we have detected the thermal torque noise at intermediate pressure, using the data to set a first preliminary upper limit on axion-like dark matter via axion-electron coupling at $20$ kHz.}

\begin{figure*}
    \centering
    \includegraphics[width=0.8\linewidth]{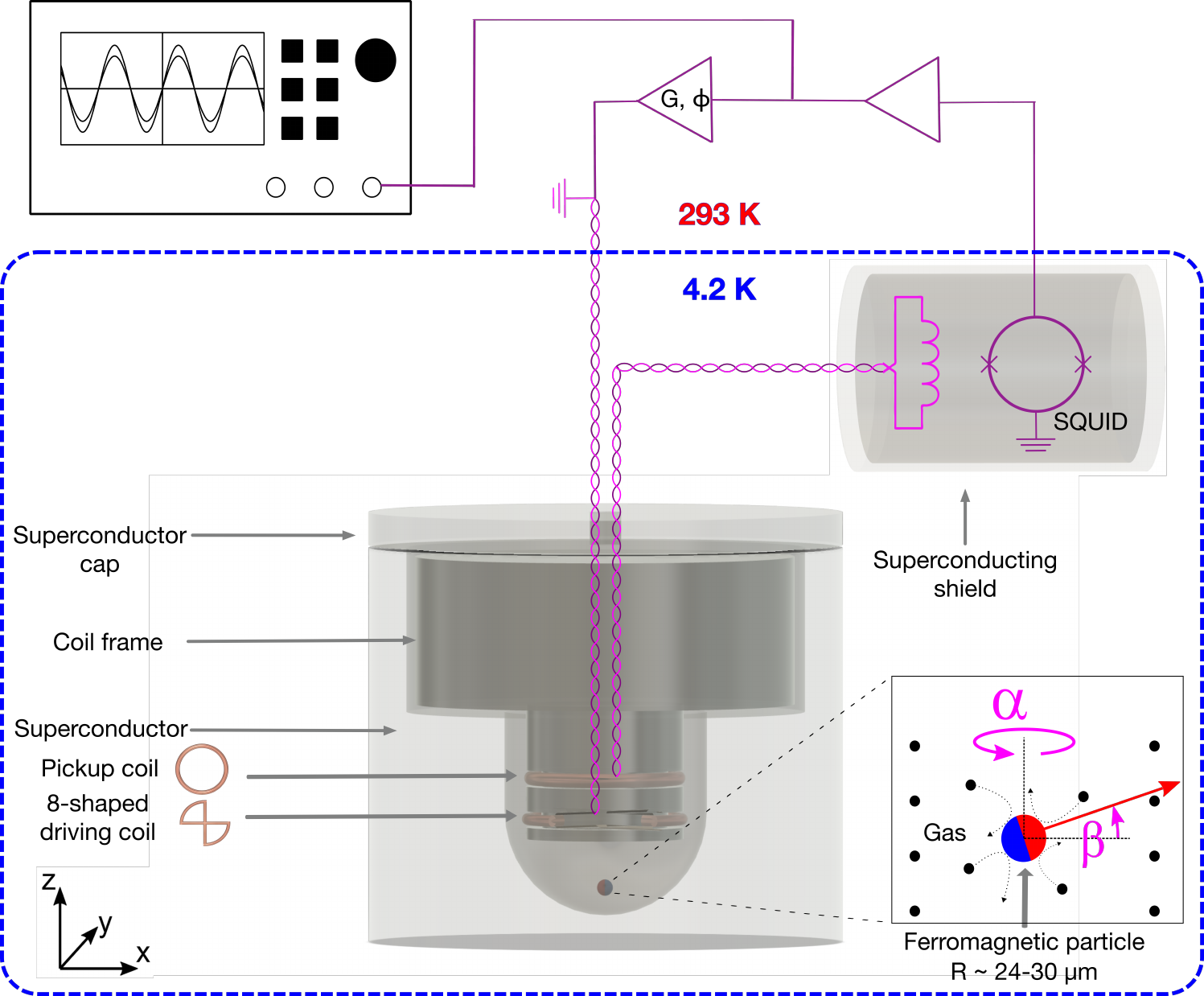}
    \caption{\textbf{Scheme of the experiment}. The setup includes a ferromagnetic particle levitated inside a superconducting lead trap, a detection system based on a dc SQUID and a pick-up coil, and a feedback circuit based on a superconducting driving coil. The trap is a cylindrical cavity with a rounded bowl-shaped bottom inside a bulk piece of lead. Trap and SQUID sit inside a vacuum chamber at $T=4.2$ K. Spinning of the magnet on the horizontal plane (rotation along $\alpha$ angle) is obtained by a synchronous driving technique. The polar angle $\beta$ is confined by the magnet-superconductor interaction. Variable gain $G$ and phase $\phi$ allow controlling magnitude and sign of the mean synchronous driving torque. The rotation signal detected by the SQUID is acquired by a digital oscilloscope. \rev{The driving coil is also used to apply a monochromatic reference field to operate the rotor in synchronization mode.}}
    \label{fig:Experimental Setup}
\end{figure*}

The experimental setup is sketched in Fig.\,\ref{fig:Experimental Setup}. Further technical details can be found in Methods.
Our rotor consists of a hard ferromagnetic microsphere made of a rare-earth alloy (the {\it magnet}), levitated in a cylindrically symmetric trap made of a type I superconductor (lead). We have performed various experiments with similar magnets with radius $R\sim 24-30$ $\mu$m. The motion of the magnet is detected by a dc SQUID flux sensor via a superconducting pick-up coil, and actuated by a driving coil. The setup is enclosed in a cryogenic vacuum chamber dipped in liquid helium at $T=4.2$ K. Before cooling, the chamber is evacuated and filled with helium gas for efficient thermalization. After cooling, we pump on the chamber to set the gas pressure to the desired value. 

The levitation of a ferromagnet above a \text{type I} superconductor has been discussed in several previous papers \cite{vinante2020ultralow, Vinante2021, Vinante2022, ahrens2025levitated, ahrens2025gyro}. In a nutshell, stable levitation relies on the Meissner repulsion between the magnetic sphere and the superconducting trap, which can be modeled as a dipole-dipole interaction between the magnetic moment of the sphere and an image dipole. At equilibrium, the magnetic dipole lies in the horizontal plane, at height $z_0$ set by the equilibrium between gravity and Meissner repulsion, while the horizontal motion is confined by the lateral surface of the lead trap (Fig. \ref{fig:Experimental Setup}). Therefore, the particle is intrinsically trapped in all translational degrees of freedom $x, y, z$. 

Following the conventions in Fig.~\ref{fig:Experimental Setup}, the rotational motion along the polar angle $\beta$ is confined by the anisotropy of the Meissner interaction, leading to a librational $\beta$ mode.
In an ideal cylindrically symmetric setup, the dipole should be free to rotate in the horizontal plane along the azimuthal angle $\alpha$. The bowl-shape is intended to suppress symmetry breaking due to tilts. In practice, however, the symmetry is broken by a residual field with a finite horizontal component, leading to another librational mode $\alpha$ \cite{vinante2020ultralow, ahrens2025levitated, ahrens2025gyro}. \rev{This field is likely related to flux trapped in the pick-up coil circuitry and wiring, and if needed it can be largely suppressed by a compensating static field \cite{ahrens2025gyro}.} The potential along the degree of freedom $\alpha$ is of type ${U=-\bm \mu \cdot \bm {B_r} =} -\mu B_r \cos \left( \alpha-\alpha_0 \right)$, where $\bm \mu $ is the magnetic dipole moment, $\bm B_r$ is the horizontal component of the residual field, and $\alpha_0$ the angle of the vector $\bm B_r$ with the direction of $x$. 
The rotational motion of the magnet on the horizontal plane will turn from librational to continuous rotation (spinning mode) if its total energy overcomes the periodic potential barrier $ E > \mu B_r$, similar to a pendulum.

To drive the magnet into spinning mode, we use a synchronous driving technique (Fig.~\ref{fig:Experimental Setup}). The angular motion signal detected by the SQUID is fed back to the driving coil, generating a driving torque which is either positive or negative, depending on the relative phase between the feedback field and the rotating magnet. This is controlled by a feedback amplifier. 
Due to the extreme sensitivity of the SQUID we use a pick-up coil with relatively weak coupling, made of a single loop. The rotational signal couples a peak-to-peak flux $\simeq  1 \, \Phi_0$ (flux quantum) in the SQUID, which is a good compromise between sensitivity and dynamic range.

The driving coil is made of a 8-shaped loop oriented along the horizontal plane. The magnetic field produced by the coil at the magnet location lies along the horizontal plane. 
To initiate the rotation, we tune the gain and phase of the feedback so as to generate a self-oscillation of the $\alpha$ librational mode, quickly leading to the free rotational regime. Once the magnet is freely spinning, further use of synchronous driving allows increasing or decreasing the rotational speed in a controlled way.

Once we have spun the magnet up to sufficiently high frequency, we study the spin-down of the free rotational motion under different conditions. The frequency is measured in real time by tracking the rotational frequency peak in the SQUID signal with a spectrum analyzer (Fig. \ref{fig:spindown}A). The equation of motion for the angular momentum under free spinning is:
\begin{equation}
    I \dot \Omega = N_d     \label{angularmomentum}
\end{equation}
where $I$ is the moment of inertia, $\Omega$ the rotational angular frequency, and $N_d$ the damping torque. For our spherical magnets we assume an isotropic moment of inertia ${I = (2/5) m R^2}$, where $R$ is the magnet radius and ${m = (4\pi/3) \rho R^3}$ the mass, with $\rho$ the mass density. Under viscous damping, the torque can be written as $N_d=-\Gamma \Omega$, with damping constant $\Gamma$. The spin-down is then described by an exponential decay of frequency:
\begin{equation}
    \Omega = \Omega_0 e^{- \gamma t}     \label{spindown}
\end{equation}
with decay rate $\gamma=\Gamma/I$. A relevant situation is when the dominant drag mechanism is gas damping. In this case, $\gamma$ is frequency-independent, and in the low-pressure free molecular regime it reads \cite{epstein1924,cavalleri2010,blakemore2020absolute}:
\begin{equation}
    \gamma = \frac{10}{\pi}\frac{P}{\rho \bar v R}  \label{gamma},
\end{equation}
where $P$ is the gas pressure and ${\bar v = \left( 8 k_B T/ \pi M_g \right) ^{1/2}}$ is the mean velocity of gas particles with mass $M_g$ at temperature $T$ according to the Maxwell-Boltzmann distribution. Notice that if the magnet is not perfectly spherical, the decay rate may depend on the rotation axis. In our experiment the asphericity is estimated at most of the order of $1\%$ (see Methods), lower than the accuracy on the estimation of $R$. The spherical approximation is thus appropriate.

According to Eq. (\ref{gamma}), $\gamma$ is a direct measurement of the pressure $P$ of the gas at the location of the magnetic rotor. It should be noted that in our experiment we have no independent access to $P$. Instead, we can independently measure the pressure in the same vacuum chamber at a different point at room temperature using a commercial Penning gauge. We label this pressure as $P_g$. As we shall discuss later, $P$ and $P_g$ may differ, due to composition and thermomolecular gradients \cite{roberts1956}.

\section{Results}

\begin{figure*}[ht]
    \centering
    \includegraphics[width = 0.75\textwidth]{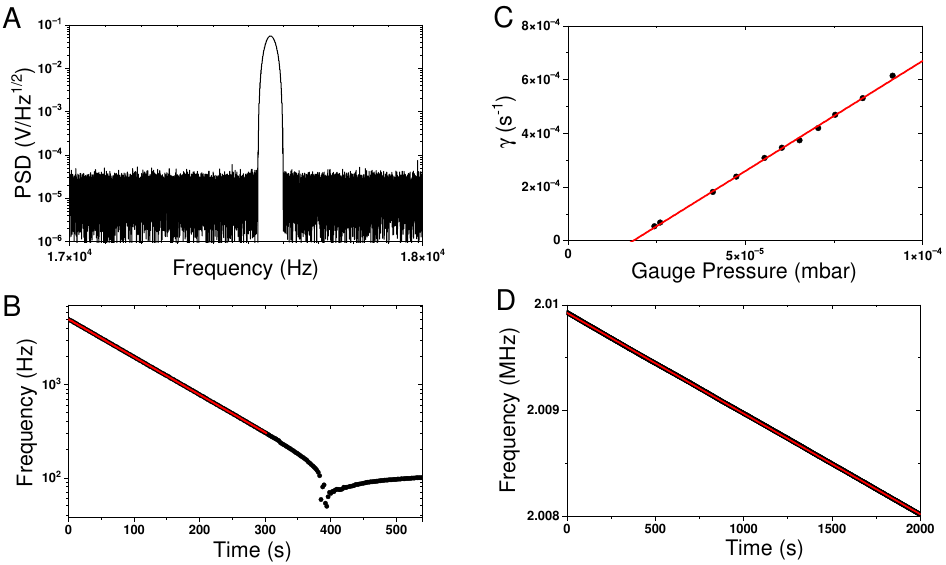}
    \caption{\textbf{Spin-down measurements.} A: Example of power spectral density (PSD) of the SQUID output signal acquired during a free spin-down. The large peak is the signal from the spinning magnet. We estimate the instantaneous spinning frequency as the frequency corresponding to the maximum of the peak in the spectrum. B: Example of full spin-down curve at the pressure $P_g=1.12 \times 10^{-3}$ mbar. Each point in the plot represents an instantaneous frequency estimated from a spectrum such as in Fig.~\ref{fig:spindown}A. The red line shows a fit of the spin-down regime with Eq. (\ref{spindown}). The abrupt interruption of the exponential spin-down at $t\sim 400$ s corresponds to the transition from free rotational motion to librational trapping, with subsequent relaxation of the frequency towards the small libration value.C: Frequency decay rate $\gamma$ as function of the pressure $P_g$ measured by the room temperature gauge. Each point in this plot is inferred from an exponential spin-down curve such as the one in Fig. \ref{fig:spindown}B. Red line is the best linear fit. D: Spin-down curve measured at a frequency $f\sim 2.01$ MHz, at the lowest effective cryogenic pressure achieved in the experiment. The $y$ axis is in logarithmic scale. The red line is the exponential fit with Eq.~(\ref{spindown}) giving $\gamma= 4.75 \times 10^{-7}$ s.}
     \label{fig:spindown}
\end{figure*}

Results on spin-down measurements are shown in Fig. \ref{fig:spindown}. In Fig.~\ref{fig:spindown}B we show a full spin-down curve for a relatively high gauge pressure $P_g=1.12 \times 10^{-3}$ mbar, where we could spin the magnet up to $5$ kHz. In the spin-down regime, we can fit with a simple exponential curve Eq.~(\ref{spindown}), finding $\gamma = 9.3 \times 10^{-3} \, \mathrm{s}^{-1}$. When the spinning frequency approaches $\sim 100$ Hz, we observe the transition from free rotational motion to trapped librational motion. In the latter regime, the motion is pendulum-like, so it features the common softening nonlinearity typical of pendulums. 

Next, we have measured the damping rate $\gamma$ as a function of the gas pressure $P_g$ measured by a room-temperature gauge, by fitting similar exponential spin-down curves at different pressures. The damping rate as a function of $P_g$ is shown in Fig.~\ref{fig:spindown}C. The dependence is linear, in agreement with Eq. (\ref{gamma}), and the red line is the best linear fit. A proper data analysis requires correcting the pressure for gas calibration and thermomolecular effects \cite{roberts1956, vinante2020ultralow}, to connect to the actual pressure $P$ of the gas at the spinning magnet location inside the cryostat. The intercept can be explained by the residual pressure of air at the location of the pressure gauge.

After pressure corrections, the slope of the fit provides a ratio $\gamma /P = \left( 11.6 \pm 0.2 \right) \, \left( \mathrm{s} \cdot \mathrm{mbar} \right)^{-1}$, in good agreement with the theoretical prediction $\gamma/P = \left( 12.0 \pm 0.2 \right) \, \left( \mathrm{s} \cdot \mathrm{mbar} \right)^{-1}$ which can be obtained from Eq. (\ref{gamma}) using the particle radius $R=(24.0 \pm 0.5) \, \mu$m  independently estimated. This confirms that the damping of the rotor provides an accurate estimate of the gas pressure at the cold side of the vacuum chamber.


We have investigated the limits in the use of the rotor as a pressure sensor by pushing to the lowest possible pressure, where other sources of damping may appear. To this end, we have cooled down the setup after evacuating the chamber, without introducing helium gas as in the normal procedure. The spin-down curve with the lowest damping is shown in Fig.~\ref{fig:spindown}D. The exponential fit yields a damping rate $\gamma = 4.75 \times 10^{-7}$, corresponding to a decay time $\tau = 1/\gamma \sim 24$ days. According to Eq.~(\ref{gamma}) this corresponds to an effective pressure $P=4 \times 10^{-8}$ mbar. This measurement shows that our levitated spinning rotor can be effective as a pressure sensor in a wide range of pressures, from $10^{-3}$ mbar to $10^{-8}$ mbar. In terms of absolute damping rate, our measured $\gamma$ is comparable to the lowest damping rate ever measured in a levitated micro or nano-object, namely a nanoparticle oscillating in a Paul trap at $1$ kHz \cite{Dania2024}. Interestingly, if we define the quality factor similarly to a resonator as $Q=\pi f /\gamma$, we obtain the enormous value $Q = 1.33 \times 10^{13}$. For an illustrative comparison, the rotational $Q$ factor of the Earth, given the decay time of $5$ billion years \cite{stephenson2016}, can be estimated as $Q = 5.6 \times 10^{12}$.  


At the lowest pressure, the synchronous driving torque is sufficient to drive the magnet to a spinning frequency well above 1 MHz. Concretely, we have achieved a maximum frequency of $2.3~\mathrm{MHz}$ for a magnet with radius $R\approx 30~\mu$m. This corresponds to a tangential speed $v=433~\mathrm{m/s}$ and a tangential acceleration $a=0.63 \times 10^{10}$ m/s$^2$. At frequencies higher than $\sim 2~\mathrm{MHz}$, close to the bandwidth of the SQUID electronics, the SQUID signal becomes significantly distorted. This sets a practical limitation to the maximum achievable rotational speed, which can be circumvented by faster electronics. Eventually, a harder limit is set by the material disintegration limit. Based on results in \cite{schuck2018} and assuming similar material breaking stress, we estimate a maximum frequency around $5.5$ MHz for a particle with radius $R=30 \, \mu$m, implying that we have operated the rotor within a factor $2$ from the breaking limit of the material. It is remarkable that, despite the enormous internal stress close to the breaking limit achieved under ultrafast rotation, the material is still maintaining ferromagnetism. This result confirms that ultrafast spinning experiments can be used to gain information on the properties of materials under extreme mechanical stress \cite{ahn2018}.
\begin{figure}[ht]
    \centering
    \includegraphics[width=1\linewidth]{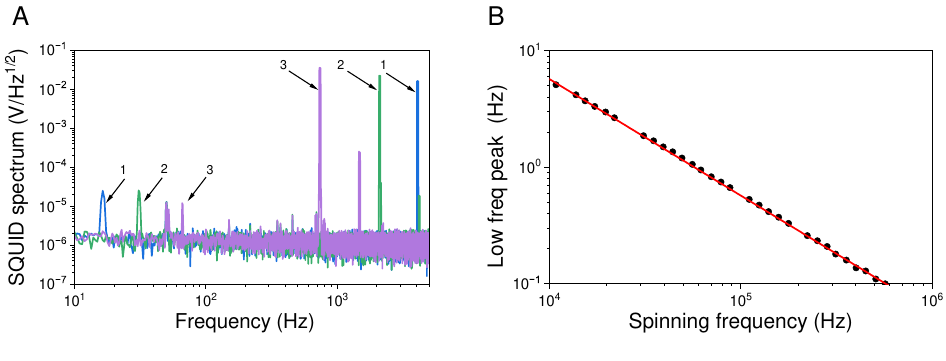}
    \caption{\textbf{Precession motion.} A: Three spectra taken from the same spin-down experiment at different times (1,2,3). The large peak at high frequency is the main signal corresponding to the free spinning $f_s$ of the magnet. A much weaker peak appears at a lower frequency $f_l$, with frequency inversely related to the spinning frequency. We attribute this peak to a precessional motion induced by the screening magnetic field produced by the superconductor above which the magnet is levitated. B: Frequency of the low frequency peak $f_l$ as a function of the spinning frequency $f_s$. The red line is a fit with the equation $f_l = f_x^2 /f_s$ predicted by our dynamical model which takes into account the superconductor field.}
    \label{fig:precession_rotation}
\end{figure}

In addition to investigating dissipation and maximum achievable frequency, we have characterized a further effect. Fig.~\ref{fig:precession_rotation}A shows a set of spectra acquired during an individual spin-down, revealing the existence of a low-frequency secondary peak, which appears to be related to the main spinning peak. We interpret the low-frequency peak as a precession of the spinning axis around the vertical axis, related to the torque induced by the image field from the superconductor, similarly to a spinning top under the effect of gravity. A complete dynamical model, discussed in the Supplementary Materials \cite{belovs2024}, predicts that the low frequency precession frequency $f_l$ is related to the spinning frequency $f_s$ by the relation $f_l=f_x^2/f_s$, where $f_x^2 \simeq 0.5 f_\beta^2$. Here, $f_\beta$ is the librational frequency along the polar angle when the magnet is not spinning. Intuitively, the connection between the precession of the spinning rotor and the librational frequency $f_\beta$ is that both are determined by the torque exerted by the image field.
Fig.~\ref{fig:precession_rotation}B shows an experimental data set with $f_l$ as a function of $f_s$. The data are well fitted by the function $f_l=f_x^2/f_s$ with the fitting parameter $f_x\simeq 240$ Hz. In this specific experiment, we measured an actual librational frequency $f_\beta=392$ Hz, predicting $f_x=f_\beta/\sqrt{2}=277$ Hz, which is close but significantly higher than the experimental value from the fit of Fig.~\ref{fig:precession_rotation}B. The slight discrepancy could be explained by the effect of the small frozen field, which breaks the rotational symmetry and determines the azimuthal trapping (i.e., the $ \alpha$ mode discussed in Fig.~\ref{fig:spindown}B). Due to this field, the measured value of $f_\beta$ is typically $\sim 10 \%$ higher than that associated with the image field and responsible for the precessional motion \cite{ahrens2025levitated}. 

\rev{Finally we have performed a first preliminary demonstration of using the rotor for ultrasensitive torque measurements. According to the fluctuation-dissipation theorem, the damping rate $\gamma$ is associated with a fundamental thermal torque noise with spectral density $S_N=4 k_B T I \gamma$. Crucially, the unique combination of low temperature and ultralow damping rate provided by our rotor ensures extremely low levels of thermal torque noise, suggesting the use of the rotor as an ultrasensitive torque sensor, with potential applications to fundamental physics. To demonstrate this possibility, we have synchronized the rotor to the oscillating magnetic field in the $xy$ plane generated via the driving coil by a local oscillator at frequency $\Omega_0$ and amplitude $B_0$. We describe the rotor with the rotation phase angle in the rotating frame $\phi(t)=\alpha(t)-\Omega_0 t$. In the presence of the oscillating field, and assuming $\dot \alpha \sim \Omega_0$, the equation of motion for $\phi$, neglecting a constant torque and fast oscillating terms, is (see Methods) :
\begin{equation}
\ddot \phi + \gamma \dot \phi + \omega_L^2 \sin \phi = \frac{N}{I},   \label{eq:synchronization}
\end{equation}
where $\omega_L^2=\mu B_0/2 I$ and $N$ is the external torque along $z$. For $\phi \ll 1$, Eq. (\ref{eq:synchronization}) describes a harmonic libration in the rotating frame. In this situation, the magnetic dipole $\bm{\mu}$ rotates synchronously with the corotating component of the external reference field, featuring small librations around it.}

\begin{figure}[ht]
\centering
\includegraphics[width=0.75\linewidth]{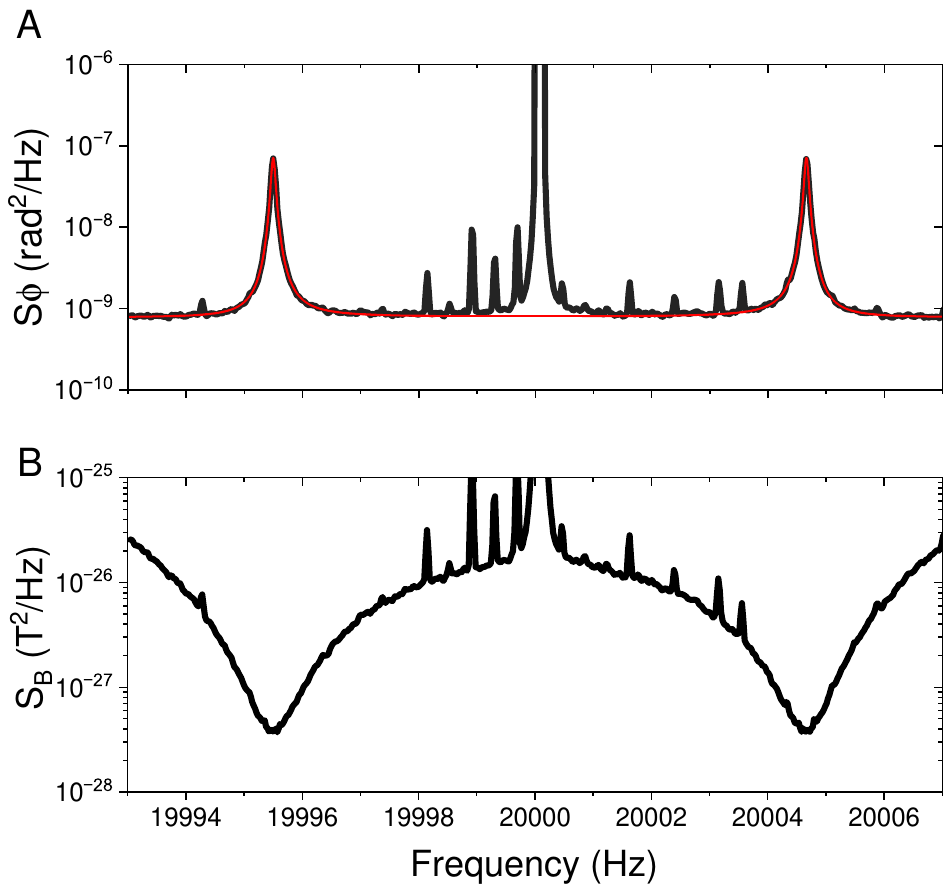}
   \caption{\rev{\textbf{Synchronized regime.} A: Spectrum of the rotor, synchronized to a reference field oscillating at frequency $\Omega_0=20$ kHz, normalized in units of phase. The huge (clipped) central peak is the main spinning signal. The two sidebands represent the $\phi$-mode libration in the rotating frame. The red curve is a fit of the two sidebands with Eq. (\ref{eq:rotornoise}). B: Magnetic field resolution, obtained by dividing the phase spectrum by the field-to-phase Lorentzian transfer function.} }
   \label{fig:rotornoise}
\end{figure}

\rev{We can trap the rotor in the synchronized regime by carefully adjusting its frequency close to the reference frequency. An example of power spectrum of a synchronized rotor with $\Omega_0=20$ kHz is shown in Fig. \ref{fig:rotornoise}A. The spectrum, normalized as a phase in radians with respect to the main rotational signal, was obtained with another particle of radius $R=(24 \pm 1) \, \mu$m at an intermediate  pressure $P\sim 4 \times 10^{-4}$ mbar, with a measured dissipation $\gamma = 4.7 \times 10^{-3} $ Hz. The large central peak at $20$ kHz is the main rotational signal, while the librational $\phi$-mode, representing a phase modulation of the main signal, appears as two sidebands. We fit both sidebands simultaneously with the function:
\begin{equation}
  S_\phi (\Omega)= A_1 + \frac{A_2}{\left[ -\left(\Omega-\Omega_0 \right)^2  + \omega_L^2 \right]^2 + \left[ \left(\Omega-\Omega_0 \right) \gamma_e \right]^2}  \label{eq:rotornoise}
\end{equation}
where the constant term $A_1$ represents the wideband noise and the double Lorentzian term describes the noise in the two sidebands (see Methods). The fit, restricted to the two sidebands yields $A_2=\left( 1.49 \pm 0.01 \right) \times 10^{-5}$ s$^{-3}$, which is in reasonable agreement with the theoretical value predicted by thermal noise $A_2=4 k_B T \gamma /I = \left( 1.1 \pm 0.2 \right) \times 10^{-5}$ s$^{-3}$, where the main uncertainty is related with the independent determination of $I$. The small excess of noise could be produced by the phase noise of the signal generator used to deliver the reference signal. We estimate that a phase noise of $-115$ dBc/Hz, that is plausible for the generator we have used, would explain the excess.}

\rev{We can now estimate the sensitivity of the synchronized rotor to an external signal produced by another in-plane weak magnetic field $\mathbf{B }$ oscillating at frequency $\Omega_0+\omega$ close to one of the two sidebands. This will cause a torque:
\begin{equation}
N \left( \omega \right) = \frac{1}{2} \mu B \left( \Omega_0 + \omega \right)   \label{torqueB}
\end{equation}
which drives the $\phi$-mode. By dividing the total observed phase noise in Figure \ref{fig:rotornoise}A by the transfer function of the applied magnetic field (see Methods), we obtain the magnetic field spectrum in Fig. \ref{fig:rotornoise}B. The minimum of about $20$ fT/Hz$^{-1/2}$ matches the best magnetic field resolution ever demonstrated with a nonrotating levitated magnet \cite{ahrens2025levitated}.}

\rev{A potential source of weak magnetic fields relevant to fundamental physics is given by axion-like dark matter \cite{Kalia2024}. According to existing models, axion-like dark matter would appear to electron spins in a ferromagnet as a quasimonochromatic classical magnetic field, with unknown frequency $f_a$ related to the axion mass $m_a$, relative coherence ${\delta f/f_a \sim 10^{-6}}$ and magnitude $B_a$ related to the dark matter density and the unknown axion-electron coupling $g_{ae}$. Therefore, a spin-based magnetometer such as our rotor can be used to probe the two-dimensional space with parameters $g_{ae}$ and $f_a$. From the data in Fig. \ref{fig:rotornoise} we do not see any candidate signal in the high sensitivity band of $\sim 2$ Hz around the sidebands and we can thus establish an upper bound on $g_{ae}$. The variance of the magnetic field $B$ in each bin of the spectrum is $\sigma^2=S_B/t$ where $t=4.2\times 10^4$ s is the integration time. Given the spectrum frequency resolution, a hypothetical axion signal would be effectively monochromatic, with a mean energy in the relevant bin $B_s^2 = \kappa B_a^2$, where $\kappa=0.33$ is an efficiency factor which includes spectral leakage and orientational efficiency, and $B_a = 4 \times 10^{-8} T \times g_{ae}$ is the mean axion field \cite{Kalia2024}. The latter relation comes from assuming the dark matter density entirely composed of axions. Setting $B_s > 3 \sigma$ as conservative detection threshold we obtain a frequency-dependent upper limit across the measured frequency window, with a best value of $g_{ae} < 1.3 \times 10^{-8}$ at the two minima in Fig. \ref{fig:rotornoise}B. This best upper limit is represented by the cyan line at 20 kHz and the zoomed inset in Fig. \ref{fig:axions}, which shows the current limits on the axion-electron coupling $g_{ae}$ on a wide frequency range. We note that there are several other peaks close to the central spinning signal that could be in principle considered as potential axion signals. We can rule out this possibility because a true axion signal would drive the $\phi$-mode showing up as two symmetric sidebands of the spinning signal. Instead, all observed peaks are either unpaired, or have a counterpart with very different amplitude. }

\section{Discussion and Prospects}

\begin{figure*}[ht]
	\centering
    \includegraphics[width=0.75 \textwidth]{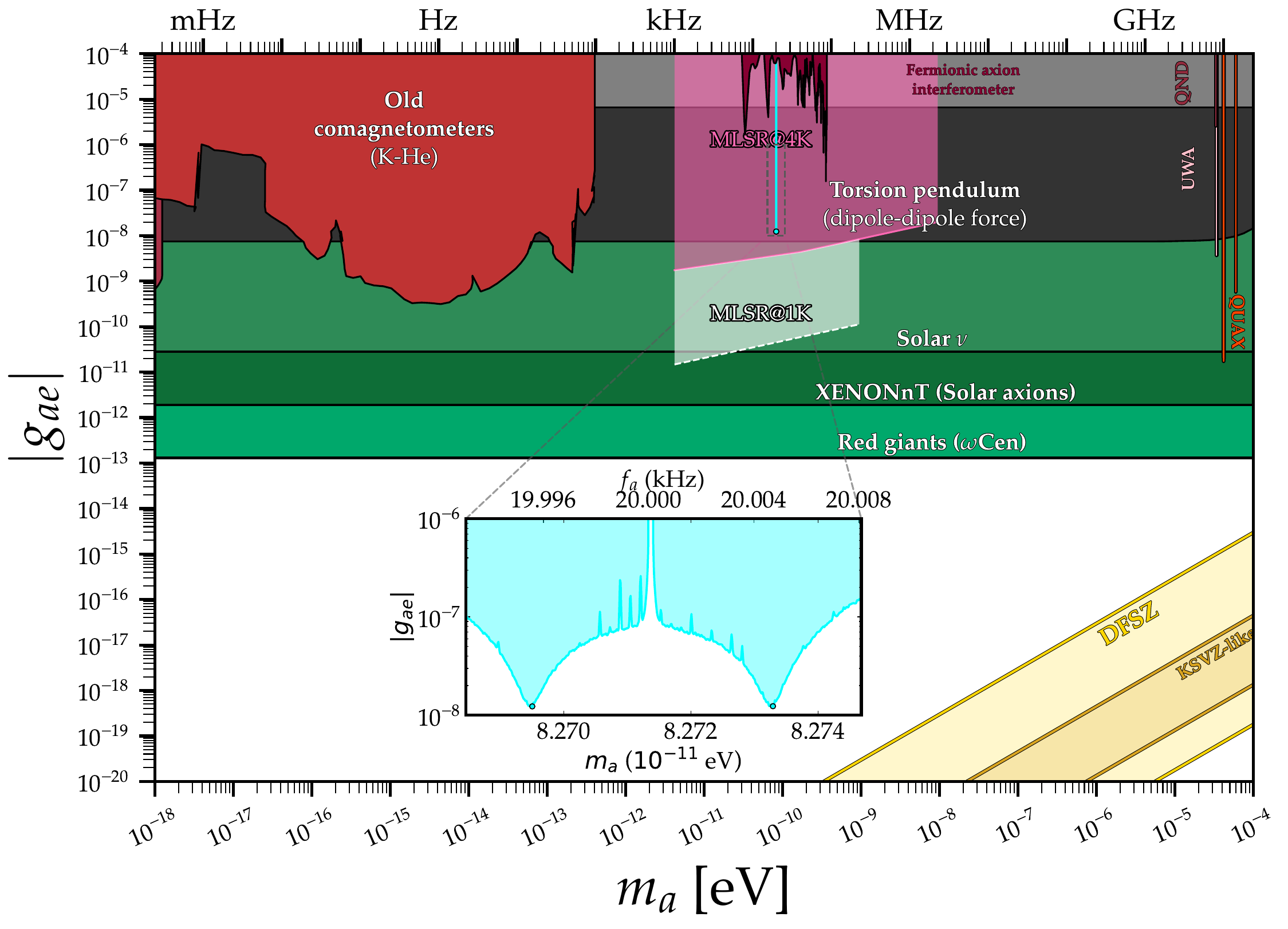}
	\caption{\rev{\textbf{Exclusion plot of the axion-electron coupling}. The cyan line labeled ``Exp'' represents the upper limit on the axion–electron coupling $g_{ae}$ obtained in the present experiment, zoomed-in in the inset. We also show the potential reach of improved versions of the experiment using the lowest achieved $\gamma$ covering frequencies from 1 kHz to 2.3 MHz (pink shaded area) and enhanced parameters in the range 1 - 230 kHz (white shaded area). The maximum frequency for the latter case is reduced by a factor 10 because we assume a sphere radius 10 times larger, reducing the breaking frequency by the same factor. We assume one can achieve the thermal noise limit even at ultralow dissipation. Other excluded areas from direct searches are shown in red, bounds from fifth-force experiments (torsion pendulums) are in gray, astrophysical bounds are in green. Light areas in the lowest right part of the plots are theoretical predictions. Figure generated using code from the AxionLimits repository \cite{AxionLimits}.}}
    \label{fig:axions}
\end{figure*}

Our experiment is, to our knowledge, the fastest magnetically levitated rotor ever demonstrated, with a maximum frequency $f_s=2.3$ MHz corresponding to 138 Mrpm. Compared to previous magnetic spinning rotors \cite{schuck2018} there are several differences. First, we levitate and drive smaller magnets, which is crucial to push the breaking limit to higher frequencies. In this respect, we note that even smaller dielectric nanoparticles with diameter below $100$ nm have been spun up to GHz frequencies using optical tweezers \cite{ahn2018,ahn2020,Zielinska2024}. Second, our setup does not require active feedback stabilization, since translational trapping is provided for free by passive Meissner-based repulsion. Lastly, conventional magnetic rotors employ asynchronous driving \cite{fremerey1972high,fremerey1985spinning,schuck2018}, which rely on inducing a large magnetic moment in the rotor via eddy currents, implying a large heat dissipation. Because of higher efficiency and much lower heating, our synchronous driving technique works well even at cryogenic temperatures, and potentially down to millikelvin temperatures. 

In fact, for the first time we can operate a spinning rotor as a cryogenic pressure sensor at $T=4.2~$K in a wide range of pressures down to $10^{-8}$ mbar. The lowest operating temperature previously reported for a spinning rotor gauge was 77 K \cite{wydra2024vima}. In general, there are very few reports of pressure sensors operating at $4.2$ K or below~\cite{raoUseExtractorGauge1993,lotzDevelopmentFieldEmitterbased2014,baglinFonctionnementDuneJauge1995}, and these are typically low-accuracy gauges that require specific calibrations, such as ionization gauges~\cite{adderleyRoadMapExtreme2008}. In contrast, a spinning rotor is considered a highly stable and accurate transfer standard for low-pressure measurements by metrological institutes \cite{fedchak2015recommended}. In the Supplementary Materials we estimate the Cramer-Rao lower bound on the precision of pressure measurement, showing that our sensor is also very sensitive and fast, with an optimal thermal-limited resolution of $3 \times 10^{-11}$~Pa/Hz$^{1/2}$ for the parameters of our system.
In addition to pressure sensors, we envision several other applications of ultrafast microrotors such as gyroscopes \cite{zeng2024}, gravimeters and magic angle spinning rotors \cite{marti2024}, and studies of material properties of materials under extreme tensile strengths \cite{schuck2018}.

\rev{As a last result, we have synchronized the rotor to an external oscillating field and measured the thermal noise in the librational $\phi$-mode in the rotating frame at an intermediate pressure, demonstrating a concrete example of using the rotor as a sensitive torque sensor. Our estimated resolution to an external magnetic field at $20$ fT/Hz$^{1/2}$ matches the best resolution ever achieved by a nonspinning levitated magnetometer of this size \cite{ahrens2025levitated} and allows us to search for axion-like dark matter with a narrowband sensitivity several orders of magnitude better than previous direct search experiments, as illustrated in Fig. \ref{fig:axions}. Our upper limit on the axion-electron coupling $g_{ae}$ is close to the best laboratory-based limits inferred from fifth-force experiments. We emphasize the very preliminary nature of this measurement and relative data analysis in the context of axion-like dark matter. A more thorough and wideband search is left to a future work, with many aspects still to be optimized by operating at larger coupling and lower dissipation. Furthermore, ultralow noise torque measurements with a synchronized rotor may require pushing the instrumental phase noise to the current technological limits. } 

\rev{Alternatively, one can operate the rotor in the free spinning mode in absence of instrumental phase noise and infer external torque by monitoring the free spindown dynamics. This unconventional approach would also offer a natural way to sweep the detection frequency.} In the Supplementary Materials we discuss quantitatively the potential reach of a sweep experiment to probe axion-like dark matter under the assumption that we can reach the thermal torque noise limit at ultralow dissipation, leading to the potential reach shown in Fig. \ref{fig:axions}. We consider the parameters measured in the current experiment (pink shaded area) as well as a set of plausibly improved parameters (white shaded area). Indeed, even lower damping and thermal noise can be expected by working at lower temperature, lower pressure or using larger magnets. This prospect is supported by recent works on diamagnetic rotors, suggesting that additional sources of dissipation relevant to oscillators and librators, such as eddy currents, can be strongly suppressed in free rotational motion \cite{kim2025}. We consider as plausible improved parameters $T=1$ K, $\gamma=10^{-8}$ Hz and $R=300 \, \mu$m, which would imply a thermal-noise-limited magnetic field noise $S_B^{1/2}=1 \times 10^{-18}\, \mathrm{T/\sqrt{Hz}}$. Our experimental bound on $g_{ae}$ would outperform existing laboratory-based experiments, and compare with model-dependent bounds derived from astrophysical considerations.

\rev{Besides the specific case of axions, an ultrasensitive rotor could be used in other fundamental physics cases. Rotors with elongated shape, featuring a large mass quadrupole moment, will enable a wider class of measurements, such as detection of high-frequency gravitational waves \cite{aggarwal2021}, mechanical tests of spontaneous collapse models \cite{vinante2016,vinante2020}, tests of quantumness versus classicality of gravity \cite{Angeli2025}. The narrowband tunable operation based on a synchronized rotor is particularly promising for two-mass experiments to measure gravity \cite{Fuchs2024} or exotic interactions \cite{Fadeev2021,Vinante2021}. Further possibilities may include tests of quantum field effects associated with fast rotational motion, such as the Zel'dovich effect \cite{Braidotti2020,Braidotti2024} and the rotational Unruh effect \cite{Lochan2020}, and investigations of fundamental quantum aspects of spins, such as the Einstein-de Haas and Barnett effect and related quantum stabilization mechanisms \cite{Einstein1915,Barnett1915,Rusconi2017,Wachter2025}. In the Supplementary Materials we derive explicit quantitative estimations of sensitivity for the specific cases of high frequency gravitational waves and test of spontaneous collapse models. }

In conclusion, we have demonstrated ultrafast and ultralow dissipation magnetic microrotors operating at cryogenic temperature, featuring spinning frequencies higher than $2$ MHz and damping lower than $10^{-6}$ Hz, demonstrating their use as a novel cryogenic pressure sensor. \rev{Furthermore, we have detected thermal torque noise at intermediate pressure, allowing us to perform a preliminary search for axion-like dark matter.} Due to the extremely low thermal noise, this system opens the way to a number of exciting opportunities in the context of fundamental and quantum science.

\section*{Material and Methods}
The magnets used in this experiment are nearly spherical microparticles made of a rare-earth alloy based on NdFeB \cite{Magnequench}, with a radius of about $24-30 \, \mu$m. In a previous work with the same type of particles \cite{ahrens2025gyro} we have measured the asphericity of many particles, defined as $\epsilon = (R_2-R_1)/(R_2+R_1)$ where $R_1$ and $R_2$ are the semi-minor and semi-major diameters. The mean value of $\epsilon$ was found to be less than $1\%$. The cylindrically symmetrical trap is made of lead, has diameter 5 mm and a bowl-shaped bottom, as shown in Fig. \ref{fig:Experimental Setup}. \rev{We choose lead as trap material because it is the type I material with the highest critical temperature, working properly at $T=4.2$ K without featuring flux vortices.} For a specific experiment, first we pick a single microsphere and magnetize it in a $10$~T NMR magnet, and subsequently we place it at the bottom of the superconducting trap. Before placing the magnet, we remove the natural tarnish layer on the lead surface using a solution of acetic acid and hydrogen peroxide. The pick-up and driving coils are made of NbTi wire with diameter $100 \, \mu$m and wound onto a 3D printed polymeric support. The coils are placed at a height of approximately 2 mm (driving) and 4 mm (pick-up) above the trap bottom. We use a commercial SQUID and a wideband SQUID electronics from Magnicon (electronics model XXF-1). 

\rev{The entire setup is enclosed in a vacuum chamber which is dipped in a liquid helium cryostat, mounted on an antivibration platform (HWL). This provides a vibrationally quiet cryogenic environment, which is crucial for high sensitivity low noise measurements.}
The pressure in the vacuum chamber is measured on the room-temperature side using a commercial Penning gauge. To connect the pressure on the room-temperature side to the pressure at the magnet location at $T=4.2$ K we have to consider that the response of the Penning gauge (calibrated to air) to helium is nonlinear, and a thermomolecular gradient is expected to appear \cite{roberts1956}. For pressures below $\sim 10^{-4}$ mbar both effects result in a simple renormalization. The Penning sensitivity to helium with respect to air is accounted for by a constant factor $5.9$. Similarly, the thermomolecular ratio between the pressure on the cryogenic side of the chamber (at temperature $T=4.2$ K), where the magnet sits, and the room temperature one (at temperature $T_{RT}\approx 295$ K) tends to the constant $(T/T_{RT})^{1/2}$ \cite{roberts1956,vinante2020ultralow}. We thus expect for helium the simple linear relation:
\begin{equation}
    P_g =\frac{1}{5.9}\left(\frac{T}{T_{RT}}\right)^{-\frac{1}{2}} P \simeq 1.42\, P.  \label{Pg}
\end{equation}
\rev{We attribute the intercept in Fig. \ref{fig:spindown}C to residual air outgassing in the room temperature side of the chamber. This is unrelated to the pressure $P$ at $4.2$ K, which is expected to be related solely to helium gas, since all other species are frozen. The intercept can in principle drift with time. For this reason, and because further removal of helium gas from the chamber below $10^{-5}$ mbar is a very slow process, we did not explore the $P-P_g$ curve for lower pressure. The measurement shown in Fig. \ref{fig:spindown}D is obtained in a different cooldown where we evacuate the chamber at room temperature without adding any helium gas. In this way we reach much lower effective pressure $P$ but we have no reliable way to compare with an independent vacuum gauge. However, the measured $\gamma$ provides an upper limit on the total amount of dissipation unrelated to the gas, and therefore allows us to conclude that the rotor can operate as a pressure sensor at least down to $10^{-8}$ mbar. }

Typically, magnets with diameter of $60 \, \mu$m or larger spontaneously levitate at the temperature $T=4.2$ K, well below the critical temperature of lead. Smaller magnets sometimes remain stuck because of electrical adhesion. In this case, we can levitate the magnet by supplying a sudden mechanical excitation, e.g. by gently hitting the cryostat with a rubber hammer.      
To spin the levitated magnet, we apply positive feedback to the librational $\alpha$ mode via one of the driving coils. This moves the magnet from trapped libration mode to spinning mode. Then we apply controlled synchronous driving by tuning the gain and phase of the feedback, so as to accelerate the rotational motion up to the desired spinning frequency.

During the spin-down, we acquire the data using an oscilloscope/spectrum analyzer (Picoscope 4000). To track the rotor frequency as a function of time, we acquire power spectra of the SQUID signal and extract the frequency corresponding to the maximum of the spinning frequency peak. It can be shown that this procedure realizes a maximum likelihood estimation of the instantaneous frequency of a rotor \cite{rife1974}. This measurement strategy is thus expected to saturate the Cramer-Rao bound, that we estimate in the SI.

An interesting consequence of spinning a magnet to very high frequency is that impurities or dust particles, that may get attached to the magnet during the preparation of the setup or during the cooldown, are efficiently expelled from the magnet surface by the strong centrifugal forces. Indeed, sometimes we observe a change of the librational frequencies by a few percent after spinning for the first time beyond some tens of kHz, likely indicating a change of the moment of inertia. After this first conditioning, the librational frequencies appear to be stable with time.

\rev{For the synchronization regime we apply a sinusoidal magnetic field using the driving coil. The signal is generated by the source of a Liquid Instruments Moku:Lab. To achieve synchronization, we carefully match the reference field to the instantaneous rotation frequency. The torque exerted by the rotating field $\mathbf{B_0}=B_0 \cos{\left( \Omega_0 t \right)}\hat x $ on the spinning magnetic dipole $\bm{\mu} = \mu \left( \cos{\alpha(t) }\hat x + \sin{\alpha}\hat y \right)$ can be calculated as the $z$-component of the vector product $\mathbf{N}=\bm{\mu} \times \mathbf{B_0}$, which is proportional to $\sin\alpha \cos(\Omega_0 t)$. If $\dot \alpha \sim \Omega_0$ one can expand the product retaining only the slowly varying term:
\begin{equation}
 N = \frac{\mu B_0}{2} \sin{\left( \alpha(t)-\Omega_0 t \right)}   \label{eq:fieldtorque}
\end{equation}
Starting from  the equation of motion in the laboratory frame:
\begin{equation}
 I \ddot \alpha + I \gamma \dot \alpha = N,
\end{equation}
inserting the expression of $N$ found above, and using the definition of the phase in the rotating frame $\phi=\alpha-\Omega_0 t$, after simple algebra and neglecting a constant term, one finally arrives to the $\phi$-mode libration Eq. (\ref{eq:synchronization}). }

\rev{The voltage signal detected by the SQUID after amplification is given by:
\begin{equation}
V=V_0 \sin \left( \Omega_0 t + \phi(t) \right) = V_0 \sin(\Omega_0 t) +  V_0 \cos(\Omega_0 t) \phi(t)
\end{equation}
where the last equality holds for $\phi\ll 1$. Thus, an oscillating phase with Fourier component $\phi (\omega)$ is translated to two voltage components with amplitude $ \phi(\Omega_0 \pm \omega)V_0/2$. The double sideband spectrum in Fig. \ref{fig:rotornoise}A is therefore normalized as a phase $\phi$ by dividing the voltage spectrum by the $V_0^2/4$ factor.
The spectrum is obtained as $1000$ averages of FFT power spectra with frequency resolution $0.0238$ Hz for a total time $t=41943$ s. To avoid spectral leakage on the sideband peaks, we apply a weak cold damping proportional to $\dot \phi$ using the Moku:Lab Phasemeter tool and a filter stage. This increases the actual dynamical bandwidth of the $\phi$-mode to $\gamma_e/2\pi \approx 0.08$ Hz, which is much larger than the FFT frequency resolution.}

\rev{According to Eq. (\ref{eq:synchronization}) for $\phi \ll 1 $ the response of $\phi$ to an external torque is:
\begin{equation}
\phi(\omega) = \chi(\omega) N(\omega)
\end{equation}
with:
\begin{equation}
\chi(\omega) = I^{-1} \left[ \left( -\omega^2 + \omega_L^2 \right) + i \omega \gamma \right]^{-1}  \label{chi}
\end{equation}
Under thermal noise driving, the power spectral density of torque is $S_N = 4 k_B T I \gamma$, and will produce a phase noise with spectrum:
\begin{equation}
  S_\phi (\omega) = |\chi(\omega)|^2 S_N 
\end{equation}
The spectrum is translated to the two sidebands of $\Omega_0$, so one has to replace $\omega$ with $\Omega_0\pm\omega$. This leads to the second term of the fitting function Eq. (\ref{eq:rotornoise}), with the expected value of the fitting parameter $A_2 = S_N/I^2$. The first term accounts for the wideband phase noise.
The reduced $\chi^2$ of the fit, extended to a bandwidth of about $\sim 2$ Hz around the two sidebands, is $\chi^2=1.22$ with $183$ degrees of freedom with a probability of $2.5\%$, meaning that the data are in reasonable agreement with Gaussian noise.
}

\rev{An external in-plane weak magnetic field $B$ will produce a torque translated in frequency by $\Omega_0$ according to Eq. (\ref{torqueB}). Therefore to produce a torque at the frequency $\omega$ of the slow $\phi$-mode libration one has to apply a field close to the two sidebands.
We obtain the magnetic field spectrum in Fig. \ref{fig:rotornoise}B by dividing the measured phase spectrum by $\mu^2/4|\chi(\Omega_0-\omega)|^2$.}

\rev{To estimate the efficiency-factor $\kappa$ we observe that the hypothetical axion signal has a coherence time $t_{coh} \approx 10^6 /(\Omega_0/2\pi) \approx 8$ s which is much smaller than the integration time. Therefore, the direction of the axion field fluctuates randomly over the measurement time. The rotor is sensitive to both in-plane components $x$ and $y$ and blind to $z$, leading to an efficiency factor $2/3$. A further loss of efficiency is provided by spectral leakage caused by windowing. We estimate the leakage based on the main spinning peak as $\sim 0.5$, leading to an efficiency factor $\kappa=0.33$.}

\rev{For the estimation of $I$ and $\mu$ we refer to the method used in Ref. \cite{ahrens2025levitated}, based on measuring the frequencies of the $z$ and $\beta$ modes of the nonspinning magnet, which provides $R=(24 \pm 1)$ $\mu$m, $I=(1.0 \pm 0.2) \times 10^{-19}$ Kg\,m$^2$ and $\mu=(4.0 \pm 0.4) \times 10^{-8}$ J/T.}


\bibliographystyle{Science}
\bibliography{references}

\section*{Acknowledgments}
A.V. thanks Matteo Carlesso for providing the code for the plot of CSL limits.

\subsection*{Funding:}

A.V., M.C., H.U., and A.C. acknowledge support from the QuantERA II Programme (project LEMAQUME) that has received funding from the European Union’s Horizon 2020 research and innovation programme under Grant Agreement No 101017733. 

A.V. and A.M. acknowledge financial support from the Italian Ministry for University and Research within the Italy-Singapore Scientific and Technological Cooperation Agreement 2023-2025.

T.W. acknowledges support from Italy-Singapore science and technology collaboration grant (R23I0IR042), Delta-Q (C230917004, Quantum Sensing). P.K.L., T.W. and J.J. acknowledges support from Competitive Research Programme (NRF-CRP30-2023-0002). J.J acknowledges the support from the LeviNet under EPSRC International Quantum Technologies Network Grant EP/W02683X/1. H.U. and M.C. acknowledge support from UKRI EPSRC grant EP/W007444/1.

\subsection*{Author Contributions}
Conceptualization: A.V., T.W., and H.U. Methodology: A.V., T. W., H.U., and A.C. Software: A.M., and M.C. Validation: A.V., T.W., and A.M. Formal analysis: A.V., T.W., A.M., A.C., and J.J. Investigation: J.J., A.M., M.C., A.V., and T.W. Writing—original draft: all authors. Writing—review and editing: A.V., T.W., A.C., A.M., M.C., and H.U. Visualization: A.M., J.J., A.V., and T.W., Supervision: A.V., T.W., P.K.L, and H.U. Funding acquisition: A.V., T.W., H.U., and P.K.L.

\subsection*{Competing interests}
The authors declare no competing interests.

\subsection*{Data, code, and material availability}
Data for this study have been deposited in the Zenodo public repository (https://doi.org/10.5281/zenodo.21650139). This study did not generate new materials or codes.


\setcounter{page}{1}

	
\begin{center}
{\Huge{Supplementary Materials for}} \\
\vspace{12pt}
\Huge{\textbf {Ultrafast Meissner-Levitated Microrotor}}
\\
\vspace{12pt}

\large{Joel K Jose \textit{et al.} \\

\vspace{24pt}

}
\end{center}

\textbf{This PDF includes:}

Supplementary Text

Figs from S1 to S7

\pagebreak

\setcounter{equation}{0}
\setcounter{figure}{0}
\renewcommand{\theequation}{S\arabic{equation}}
\renewcommand{\thefigure}{S\arabic{figure}}

\section*{Highest measured spinning frequency}

\begin{figure}[h]
	\centering
	\includegraphics[width=0.7\textwidth]{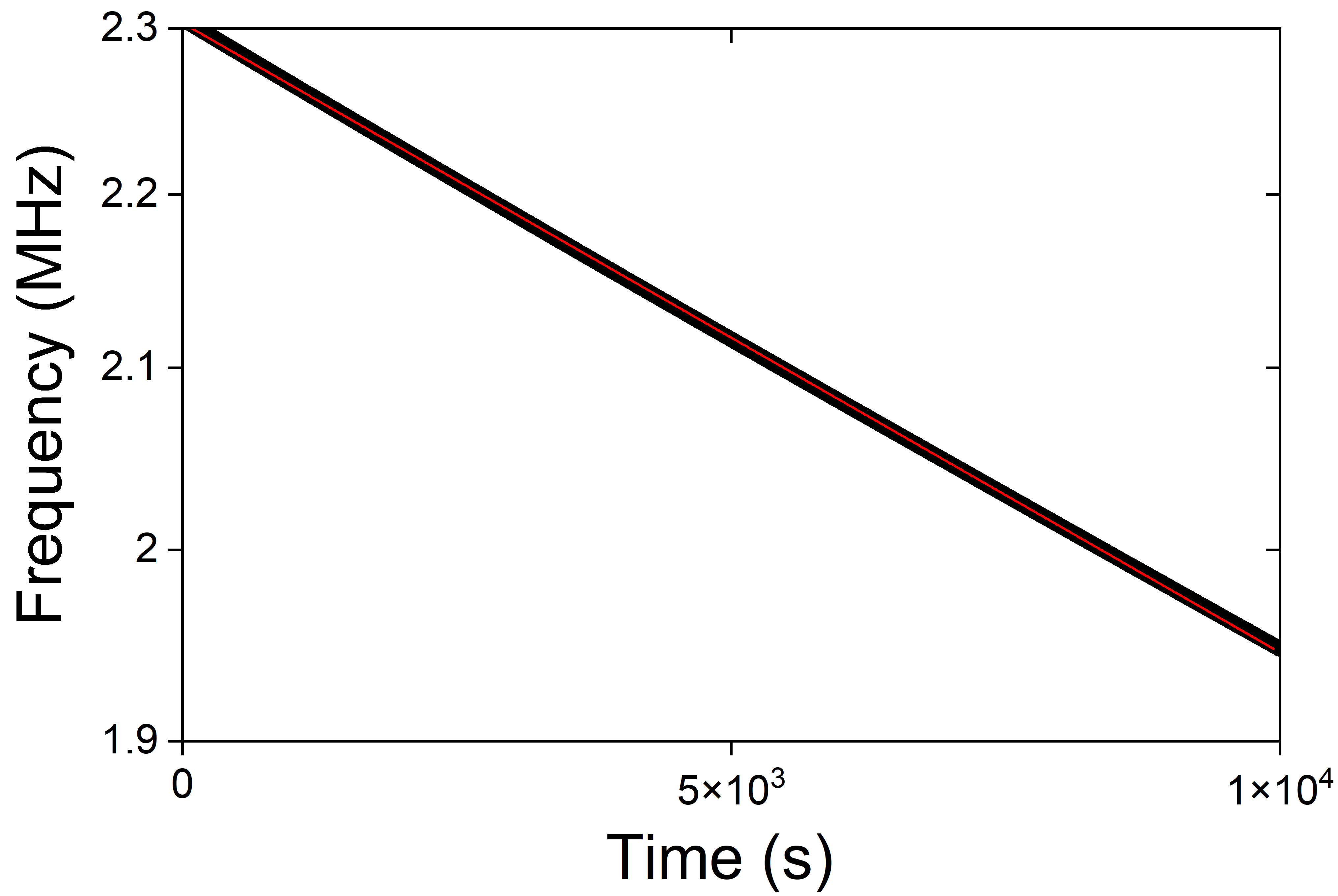}
	\caption{\textbf{Highest measured spinning frequency.} Spin-down curve of a magnet with radius $R \simeq 30 \, \mu$m, starting from an initial frequency, after driving up, of $2.3$ MHz. This is the highest frequency recorded in our series of experiments. The damping rate for the spin-down is $\gamma=1.69 \times 10^{-5}$ s.}
	\label{fig:1}
\end{figure}

In Fig.\,\ref{fig:1} we show a spin-down curve where the highest spinning frequency measured in our experiment $f_s=2.3$ MHz. For this particular measurement we used a magnet with radius of $30 \, \mu$m. The spin-down curve can still be fitted by an exponential curve, in this case with $\gamma=1.69 \times 10^{-5}$ s.

At this pressure, we could have reached an even higher frequency, but the driving time was quite long, on the order of several hours. The distortion resulting from the finite bandwidth of the SQUID electronics was significant but not yet a hard limiting factor. 
A harder limit is expected to be set by the material breaking limit.  According to the analysis in Ref. \cite{schuck2018}, material breaking is achieved when the maximum stress in the center, given by:
\begin{equation}
 \sigma = K \rho \Omega^2 R^2 
\end{equation}
overcomes the breaking stress $\sigma_{\mathrm{max}}$.
This implies that the maximum frequency scales as:
\begin{equation}
 \Omega_{\mathrm{max}} = \frac{1}{R} \sqrt{\frac{\sigma_{\mathrm{max}}}{K \rho}},
\end{equation}
where $K$ is a geometrical factor.
Thus, $\Omega_{\mathrm{max}}$ is inversely proportional to the radius $R$, corresponding to a radius-independent tangential velocity.

For the stainless steel spheres used in Ref. \cite{schuck2018}, the geometric constant $K=0.398$ was estimated and the particles with $R=250 \, \mu$m were found to explode at $\Omega/2\pi \simeq 660$ kHz. Since our particles have very similar density and geometry, if we assume a similar breaking stress, for our particle radius $R\approx 30\,\mu$m we expect disintegration to occur at $\Omega/2\pi \approx 5.5$ MHz. Prudentially, we decided to stay less than a factor $2$ below this threshold, to avoid the risk of explosion, which could damage the lead trap.

Another related effect expected at high rotational speed is an elastic deformation of the sphere into an oblate spheroid which can be calculated using standard formulas from the theory of elasticity \cite{Fremerey1973}.
The equatorial deformation is given by:
\begin{equation}
    \frac{\delta R}{R} = \frac{8- \nu - 5 \nu^2}{5 \left( 7+ 5 \nu \right)} \left(\frac{v}{v_s}\right)^2
\end{equation}
where $v = \Omega R$ is the tangential velocity at the equator, $v_s = (E/\rho)^{1/2}$ with $E$ Young modulus is the sound velocity, and $\nu$ is the Poisson modulus. Using standard NdFeB as reference material, $E\approx 160$ GPa, $\nu=0.24$, for the maximum rotational frequency $\Omega/2\pi=2.3$ MHz at radius $R=30$ $\mu$m we estimate $v=433$ m/s, $v/v_s=0.09$ and $\delta R/R = 1.6 \times 10^{-3}$. The effect is sizeable but still small. For higher frequency approaching the breaking limit one expects a comparable relative apparent reduction of the damping rate due to the dependence of the moment of inertia on frequency \cite{Fremerey1973}.


\section*{Full dynamical model and precession motion}

In this section, we develop a full model of the magnet spinning above a superconductor, and then discuss the appearance of a slow precession motion as observed in the experiment.
The equations of motion of the magnetic particle, taking into account gyromagnetism, are based on earlier works \cite{belovs2024}. In the case of a particle levitated above a superconductor, it is necessary to consider the currents in the superconductor arising due to the Meissner effect. Their action on the particle is described by the potential energy:
\begin{equation}
U=\frac{\mu^{2}}{2(2z)^{3}}(1+\cos^{2}{(\vartheta)}) ,
\label{Eq:1}
\end{equation}
where $\mu$ is the magnetic moment of the particle, $z$ is its distance above the superconductor, $\vartheta$ is the angle between the magnetic moment of the particle and the normal to the superconductor. We assume for simplicity that the magnet levitates above an infinite plane. 

Due to the dependence of the potential energy on the orientation of the dipole, there is a torque on the particle $\vec{N}$ that can be expressed as
\begin{equation}
\vec{N}=\mu H_{a}(\vec{e}\cdot\vec{e}_{z})[\vec{e}_{z}\times\vec{e}]
\label{Eq:2}
\end{equation}
where $\vec{e}$ is the unit vector along the magnetic moment $\vec{\mu}=\mu\vec{e}$, $H_{a}=\mu/(2z)^{3}$.

In \cite{belovs2024} the equations of motion of the particle are derived considering the conservation of energy and total angular momentum. In the dissipationless case for the spherical particle and in the absence of an external field, they read:
\begin{equation}
I\frac{d\vec{\Omega}}{dt}+\frac{1}{\gamma_0}\frac{d\vec{\mu}}{dt}=\vec{N}
\label{Eq:3}
\end{equation}
\begin{equation}
\frac{d\vec{\mu}}{dt}=[\vec{\Omega}\times\vec{\mu}]
\label{Eq:4},
\end{equation}
where $\gamma_0<0$ is the gyromagnetic ratio, $I$ is the momentum of inertia of the particle. The rotational drag on the particle may be accounted for by adding in the right part of Eq.~(\ref{Eq:3}) a term $-\alpha_{p}\vec{\Omega}$. In the conditions of experimental setup, its influence is very small and can be neglected.

The unit vector $\vec{e}$ is defined by the relation $\vec{e}=(\sin{(\vartheta)}\sin{(\varphi)},-\sin{(\vartheta)}\cos{(\varphi)},\cos{(\vartheta)})$. Note that here we use for convenience the standard spherical coordinate system, which differs slightly from the one in Fig.~1, in that $\beta = \pi/2- \vartheta$, while $\alpha=\varphi$. The frequency of libration of the particle around the axis parallel to the superconductor, neglecting the small contribution due to the gyromagnetic effects, reads $\omega^{2}_{\beta}=\mu H_{a}/I$. It turns out to be convenient to take as the scaling factor the angular velocity of the particle $\Omega_{p}=\omega_{\beta}$. Then $\tilde{t}=\Omega_{p}t,\vec{\Omega}=\Omega_{p}\tilde{\vec{\Omega}}$ (the tildes are omitted in the following). Using scalings Eq.~(\ref{Eq:3}) gives
\begin{equation}
\Omega^{2}_{p}\frac{d\vec{\Omega}}{dt}-\frac{\mu}{|\gamma_0|I}\Omega_{p}\frac{d\vec{e}}{dt}=\Omega_{p}^{2}(\vec{e}_{z}\cdot\vec{e})[\vec{e}_{z}\times\vec{e}]
\label{Eq:5}
\end{equation}
The second equation does not change in dimensionless form:
\begin{equation}
\frac{d\vec{e}}{dt}=[\vec{\Omega}\times\vec{e}].
\label{Eq:6}
\end{equation}
As a result, the small parameter $\varepsilon=\omega_{E}/\Omega_{p}$ appears, where 
$\omega_{E}=\mu/(|\gamma_0|I)$ is the characteristic Einstein-de Haas frequency, showing the importance of gyromagnetic effect. In the present experiment $\varepsilon=10^{-3}$. We note that for the effects considered in the following, a nonzero value of the parameter $\varepsilon$ is essential. 

Integration of the set of nonlinear equations 
\begin{equation}
\frac{d\vec{\Omega}}{dt}-\varepsilon\frac{d\vec{e}}{dt}=(\vec{e}_{z}\cdot\vec{e})[\vec{e}_{z}\times\vec{e}]
\label{Eq:7}
\end{equation}
\begin{equation}
\frac{d\vec{e}}{dt}=[\vec{\Omega}\times\vec{e}]
\label{Eq:8}
\end{equation}
is possible due to integrals of motion. It is easy to see that $\Omega_{\xi}=\vec{\Omega}\cdot\vec{e}$ is an integral of motion. We now set the value $\Omega_{\xi}=0$, which corresponds to the unit vector $\vec{e}$ orthogonal to the angular velocity $\vec{\Omega}$. This situation is supposedly close to the experimental one. Introducing the angle of rotation $\psi$ around $\vec{e}$  this condition gives $\dot{\psi}=-\dot{\varphi}\cos{(\vartheta)}$. The components of angular velocity in the laboratory frame gives
\begin{eqnarray}
\Omega_{x}=\dot{\vartheta}\cos{(\varphi)}+\dot{\psi}\sin{(\vartheta)}\sin{(\varphi)} \\
\Omega_{y}=\dot{\vartheta}\sin{(\varphi)}-\dot{\psi}\sin{(\vartheta)}\cos{(\varphi)} \\
\Omega_{z}=\dot{\varphi}+\dot{\psi}\cos{(\vartheta)}
\end{eqnarray}
As a result, the angular velocity reads:
\begin{equation}
\vec{\Omega}=(\cos{(\varphi)}\dot{\vartheta}-\cos{(\vartheta)}\sin{(\vartheta)}\sin{(\varphi)}\dot{\varphi},
\sin{(\varphi)}\dot{\vartheta}+\cos{(\vartheta)}\sin{(\vartheta)}\cos{(\varphi)}\dot{\varphi},\sin^{2}{(\vartheta)}\dot{\varphi}).
\end{equation}
Calculating $\frac{d\Omega_{x}}{dt}$ and $\frac{d\Omega_{y}}{dt}$ and their combination $\frac{d\Omega_{x}}{dt}\cos{(\varphi)}+\frac{d\Omega_{y}}{dt}\sin{(\varphi)}$
we have:
$
\frac{d\Omega_{x}}{dt}\cos{(\varphi)}+\frac{d\Omega_{y}}{dt}\sin{(\varphi)}=\varepsilon\Bigl(\frac{de_{x}}{dt}\cos{(\varphi)}+\frac{de_{y}}{dt}\sin{(\varphi)}\Bigr)-e_{z}e_{y}\cos{(\varphi)}+e_{z}e_{x}\sin{(\varphi)}
$
This relation gives:
\begin{equation}
\ddot{\vartheta}-\dot{\varphi}^{2}\cos{(\vartheta)}\sin{(\vartheta)}=\varepsilon\sin{(\vartheta)}\dot{\varphi}+\cos{(\vartheta)}\sin{(\vartheta)}.
\label{Eq:9}
\end{equation}
The z-component of Eq.~(\ref{Eq:7}) gives:
\begin{equation}
\frac{d}{dt}\Bigl(\sin^{2}{(\vartheta)}\dot{\varphi}-\varepsilon\cos{(\vartheta)}\Bigr)=0.
\label{Eq:10}
\end{equation}
We set the value of the integral equal to $\Omega_{0}$. Then:
\begin{equation}
\dot{\varphi}=\frac{\Omega_{0}+\varepsilon\cos{(\vartheta)}}{\sin^{2}{(\vartheta)}}.
\label{Eq:11}
\end{equation}
As a result, we have:
\begin{equation}
\ddot{\vartheta}-\frac{(\Omega_{0}+\varepsilon\cos{(\vartheta)})(\Omega_{0}\cos{(\vartheta)}+\varepsilon)}{\sin^{3}{(\vartheta)}}-\cos{(\vartheta)}\sin{(\vartheta)}=0
\label{Eq:12}
\end{equation}
which, by introducing the effective potential energy $U$, may be put in the form:
\begin{equation}
\ddot{\vartheta}+\frac{\partial U}{\partial \vartheta}=0
\label{Eq:13}
\end{equation}
where
\begin{equation}
U=\frac{1}{2}\frac{\varepsilon^{2}+\Omega^{2}_{0}+2\varepsilon\Omega_{0}\cos{(\vartheta)}}{\sin^{2}{(\vartheta)}}+\frac{1}{2}\cos^{2}{(\vartheta)}.
\label{Eq:14}
\end{equation}
Eq.~(\ref{Eq:13}) gives the integral of motion
\begin{equation}
\frac{1}{2}\dot{\vartheta}^{2}+U=const.
\label{Eq:15}
\end{equation}

There is an alternative way to derive the relation (\ref{Eq:15}) and the value of the integral. Multiplying Eq.~(\ref{Eq:7}) by $\vec{\Omega}$ we have:
\begin{equation}
\frac{d}{dt}\Bigl(\frac{1}{2}\vec{\Omega}^{2}+\frac{1}{2}\cos^{2}{(\vartheta)}\Bigr)=0.
\label{Eq:16}
\end{equation}
The rotational energy reads:
\begin{equation}
\frac{1}{2}\vec{\Omega}^{2}=\frac{1}{2}\Bigl(\dot{\vartheta}^{2}+\dot{\varphi}^{2}\sin^{2}{(\vartheta)}\Bigr)
\label{Eq:17}
\end{equation}
leading to the integral of motion:
\begin{equation}
\frac{1}{2}\dot{\vartheta}^{2}+\frac{1}{2}\dot{\varphi}^{2}\sin^{2}{(\vartheta)}+\frac{1}{2}\cos^{2}{(\vartheta)}=E_{n}.
\label{Eq:18}
\end{equation}
As a result we obtain:
\begin{equation}
\frac{1}{2}\dot{\vartheta}^{2}+U-\frac{1}{2}\varepsilon^{2}=E_{n}
\label{Eq:19}
\end{equation}

Let us illustrate these results by numerical examples. In the case of initial conditions $\vec{\Omega}=(0,\Omega_{y},\Omega_{0}),\vartheta(0)=\pi/2$ we have $E_{n}=(\Omega_{y}^{2}+\Omega^{2}_{0})/2$. The interval of nutation angle $\vartheta\in[\vartheta_{1},\vartheta_{2}]$ is found from $U(\vartheta_{1,2})=E_{n}+\varepsilon^{2}/2 $.
\begin{figure}
	\centering
	\includegraphics[width=0.7\textwidth]{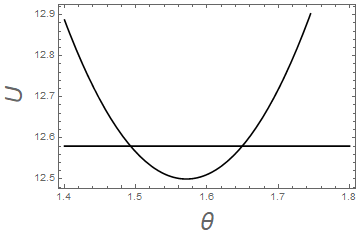}
	\caption{\textbf{Effective nutation potential.} Potential energy in dependence on the angle $\vartheta$ and the value of the energy $E_{n1}=12.58$ determined by the initial conditions $\vec{\Omega}(0)=(0,0.4,5);\vartheta(0)=\pi/2$. $\varepsilon=10^{-3}$}
	\label{fig:2}
\end{figure}
A numerical example for $\varepsilon=10^{-3}$ and $\vec{\Omega}(0)=(0,0.4,5);\vartheta(0)=\pi/2$ is illustrated in Fig.~\ref{fig:2}, where the potential energy and the integration constant $E_{n1}=E_{n}+\varepsilon^{2}/2=12.58$ are shown. The period:
\begin{equation}
T=\int^{\vartheta_{2}}_{\vartheta_{1}}d\vartheta\sqrt{\frac{2}{E_{n1}-U(\vartheta)}},
\end{equation}
for the given numerical values of the parameters gives $T=1.23$, is in good agreement with the value of the period calculated from the Fourier spectrum of $e_{z}(t)$.

\begin{figure}
	\includegraphics[width=0.5\textwidth]{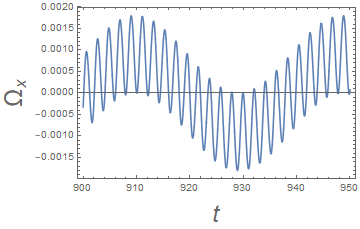}
        \includegraphics[width=0.5\textwidth]{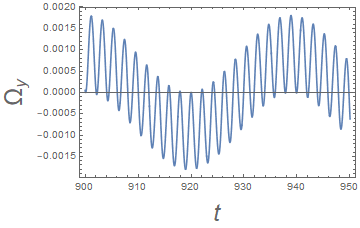}
	\caption{\textbf{Numerical simulations of fast and slow oscillations.} Time dependence of $\Omega_{x}(t)$ and $\Omega_{y}(t)$, which reveal the superposition of two oscillations. The $\pi/2$ phase shift between the low frequency oscillation on $\Omega_x$ and $\Omega_y$ shows that the oscillation corresponds to a precession of the spinning plane. Here the parameters are $\vec{\Omega}(0)=(0,0,\Omega_{0})$,$\vec{e}(0)=(1,0,0)$, $\Omega_{0}=3$ and $\varepsilon=10^{-3}$.  }
	\label{fig:3}
\end{figure}

\begin{figure}
	\centering
	\includegraphics[width=0.7\textwidth]{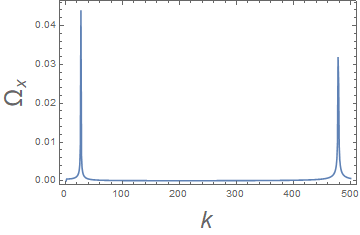}
	\caption{\textbf{Spectrum of fast and slow oscillations.} Fourier spectrum of data shown in Fig.~\ref{fig:3}.}
	\label{fig:4}
\end{figure}

The particle dynamics with initially enforced rotation is studied by numerically solving equations (\ref{Eq:7},\ref{Eq:8}). We take the following set of initial conditions $\vec{\Omega}(0)=(0,0,\Omega_{0})$ and $\vec{e}(0)=(1,0,0)$, so that $\vec{\Omega}\cdot\vec{e}=0$ as previously assumed. In the considered range of $\Omega_{0}$ a superposition of two rotation frequencies can be seen as shown in the time dependence of $\Omega_{x}(t)$ and $\Omega_{y}(t)$ in Fig.~\ref{fig:3} $(\Omega_{0}=3)$. This is in qualitative agreement with the experimental observation of a secondary low-frequency peak. The highest frequency $f_{s}$ corresponds to the spinning of the particle with period $2\pi/\Omega_{0}$. This could be seen from the time dependence of $e_{x}(t),e_{y}(t)$. 
The low-frequency oscillation can be interpreted as a slow precession of the spinning plane around the vertical axis, as apparent from the $\pi/2$ phase shift between the oscillations in $\Omega_x$ and $\Omega_y$. This behaviour resembles closely that of a spinning top under the effect of gravity.
\begin{figure}
	\centering
	\includegraphics[width=0.7\textwidth]{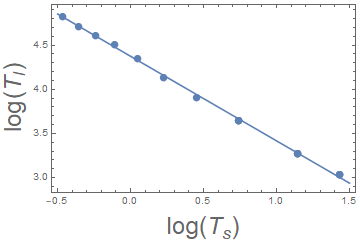}
	\caption{\textbf{Relation between the periods of fast and slow oscillations.} Periods of slow $T_{l}$ and fast $T_{s}$ modes for different values of $\Omega_{0}$. Parameters are the same as in Fig.~\ref{fig:2}.}
	\label{fig:5}
\end{figure}
The superposition of two frequencies can be explicitly seen in the Fourier spectrum (Fig.~\ref{fig:4}). The periods of slow $T_{l}$ and fast $T_{s}$ modes are numerically evaluated for different initial frequencies in Fig.~\ref{fig:5}. A linear fit in the range shown in Fig.~\ref{fig:5} gives $\ln{(T_{l})}=\ln{(C)}-\alpha\ln{(T_{s})}$ where $\ln{(C)}=4.3744$ and $\alpha=0.96$. For smaller $T_{s}$, $\alpha$ is closer to 1. In dimensional units and setting $\alpha=1$ we obtain:
\begin{equation}
f_{l}f_{s} \simeq 0.5 f_{\beta}^{2},
\label{Eq:20a}
\end{equation}
where $f_\beta=\omega_\beta / 2\pi$.
The experimental data are in reasonable agreement with this relation, as discussed in the main text.

\section*{Fundamental Sensitivity Limit of the MLSR Gauge}

We derive the fundamental pressure sensitivity of a Meissner levitated spinning rotor (MLSR) gauge. The pressure is inferred from the spin-down rate \(\gamma\), and the corresponding Cramér--Rao lower bound (CRLB) is evaluated including both SQUID readout noise and thermal torque noise. For the readout contribution we use Tretter’s result for frequency estimation of a noisy sinusoid~\cite{tretter1985estimating}, adapted to the present problem.

\subsection*{Dynamics and observable}

We consider a single rotational degree of freedom of a levitated micromagnet. The angular velocity decays exponentially,
\begin{equation}
\Omega(t)=\Omega_0 e^{-t/\tau},
\qquad
\gamma \equiv \tau^{-1},
\end{equation}
where \(\Omega_0\) is the initial angular velocity and \(\tau\) the damping time constant.

The accumulated phase is obtained by integrating the angular velocity,
\begin{equation}
\alpha(t)=\alpha_0+\int_0^t \Omega(s)\,\mathrm ds
=\alpha_0+\Omega_0\tau\bigl(1-e^{-t/\tau}\bigr),
\end{equation}
where \(\alpha_0\) is the initial phase.

Defining the spin frequency \(f(t)=\Omega(t)/(2\pi)\), it is convenient to work with the logarithmic observable
\begin{equation}
x(t)\equiv \ln f(t),
\end{equation}
which linearizes the exponential decay. In the absence of noise,
\begin{equation}
\frac{\mathrm d x}{\mathrm dt}=-\gamma,
\end{equation}
so the damping rate \(\gamma\) appears as the slope of \(x(t)\) versus time. Estimating \(\gamma\) therefore reduces to a linear regression problem.

\subsection*{SQUID readout noise}

The SQUID detects the periodic magnetic signal produced by the rotating magnet. The output is modeled as
\begin{equation}
y(t)=A\sin\alpha(t)+n(t),
\end{equation}
where \(A\) is the signal amplitude and \(n(t)\) additive Gaussian readout noise.

The intrinsic SQUID noise is characterized by the one-sided flux-noise power spectral density (PSD) \(S_\Phi(\omega)\). At the quantum limit, the flux noise is bounded by
\begin{equation}
S_\Phi^{1/2} \gtrsim \sqrt{2\hbar L},
\end{equation}
where \(L\) is the SQUID loop inductance. In practice, the actual noise level may exceed this bound and can exhibit frequency dependence.

After quadrature demodulation and phase unwrapping, the signal can be expressed in complex form,
\begin{equation}
y(t_k)=A e^{i[\alpha(t_k)+\phi_n(t_k)]},
\end{equation}
where \(\phi_n(t)\) represents phase noise with PSD
\begin{equation}
S_\phi(\omega)=\left|K_{\Phi\to\phi}(\omega)\right|^2 S_\Phi(\omega),
\end{equation}
and \(K_{\Phi\to\phi}\) is the flux-to-phase responsivity. Sampling at \(t_k=k\Delta\) yields the discrete-time phase measurement
\begin{equation}
\tilde{\alpha}_k=\alpha(t_k)+\nu_k,
\qquad k=0,\dots,N_s-1,
\end{equation}
where \(\nu_k\) is a stochastic phase-noise process.

Following Tretter~\cite{tretter1985estimating}, the variance of the frequency estimator \(\hat{\Omega}\) depends on the phase-noise PSD,
\begin{equation}
\mathrm{Var}(\hat{\Omega})
=
\frac{12}{t_m^2 N_s(N_s^2-1)}
\int_{-\infty}^{\infty} S_\phi(\omega)\,|G(\omega)|^2\,\mathrm d\omega,
\qquad t_m=N_s\Delta,
\end{equation}
where \(G(\omega)\) is the Fourier transform of the regression weighting function. This expression explicitly shows how the spectral properties of the readout noise determine the estimator variance.

In the limit of large sample number \(N_s\gg1\), and for white phase noise,
\begin{equation}
\nu_k\sim \mathcal N(0,\sigma_\nu^2),
\end{equation}
this reduces to
\begin{equation}
\mathrm{Var}(\hat{\Omega})
\approx
\frac{12\sigma_\nu^2}{f_s t_m^3},
\qquad
f_s=\Delta^{-1}.
\end{equation}

The effective measurement noise variance \(\sigma_v^2\) in the observable \(x_k\) is determined by the SQUID flux-noise PSD. For white noise and sampling interval \(\Delta\),
\begin{equation}
\sigma_v^2 \approx \frac{K_{\Phi\to x}^2 S_\Phi}{2\Delta},
\end{equation}
so that the readout-limited sensitivity ultimately scales with \(S_\Phi^{1/2}\).

In the MLSR gauge, we estimate the decay rate \(\gamma\) from
\begin{equation}
x_k \equiv x(t_k),
\qquad
z_k=x_k+v_k,
\end{equation}
with
\begin{equation}
x_k\approx x_0-\gamma t_k,
\end{equation}
and measurement noise \(v_k\sim\mathcal N(0,\sigma_v^2)\). The CRLB for \(\gamma\) is therefore
\begin{equation}
\mathrm{Var}(\hat\gamma)_{\mathrm{SQUID}}
=
\frac{12\sigma_v^2}{f_s t_m^3},
\end{equation}
which represents the SQUID readout-limited contribution to the sensitivity.

\subsubsection*{Thermal torque noise}

Thermal fluctuations in the residual gas produce stochastic torque on the rotor. The rotational dynamics are described by
\begin{equation}
I\,\mathrm d\Omega(t)
=
-I\gamma\,\Omega(t)\,\mathrm dt
+\sqrt{2I\gamma k_B T}\,\mathrm dW_t,
\end{equation}
an Ornstein--Uhlenbeck process consistent with the fluctuation--dissipation theorem.

Linearizing \(x(t)=\ln\Omega(t)\) about \(\Omega_0\) gives
\begin{equation}
\mathrm d x(t)=-\gamma\,\mathrm dt+\sqrt{\mathcal Q}\,\mathrm dW_t,
\qquad
\mathcal Q=\frac{2\gamma k_B T}{I\Omega_0^2}.
\end{equation}
Thus thermal torque appears as process noise in the linearized dynamics.

\subsection*{Discrete-time model and exact CRLB}

Sampling at \(t_k=k\Delta\) gives
\begin{align}
x_{k+1} &= x_k-\gamma\Delta+w_k,
& w_k &\sim \mathcal N(0,\mathcal Q\Delta),\\
z_k &= x_k+v_k,
& v_k &\sim \mathcal N(0,\sigma_v^2).
\end{align}

The measurement vector \(\mathbf z=[z_0,\dots,z_{N_s-1}]^\top\) has mean
\begin{equation}
\mu(\eta,\gamma)=\eta \mathbf 1-\gamma \mathbf t,
\end{equation}
with
\begin{equation}
\mathbf t=[0,\Delta,\dots,(N_s-1)\Delta]^\top.
\end{equation}

The covariance matrix is
\begin{equation}
\Sigma_{ij}=\mathcal Q\Delta\,\min(i,j)+\sigma_v^2\delta_{ij}.
\end{equation}

Eliminating the nuisance parameter \(\eta\), the CRLB for \(\gamma\) is
\begin{equation}
\mathrm{Var}(\hat\gamma)\ge
\Bigl[\mathbf t_\perp^{\!\top}\Sigma^{-1}\mathbf t_\perp\Bigr]^{-1},
\end{equation}
with
\begin{equation}
\mathbf t_\perp
=
\mathbf t
-
\frac{\mathbf 1^\top \Sigma^{-1}\mathbf t}{\mathbf 1^\top \Sigma^{-1}\mathbf 1}\,\mathbf 1.
\end{equation}

\paragraph{Readout-limited (\(\sigma_v^2\gg \mathcal Q t_m\)).}
\begin{equation}
\mathrm{Var}(\hat\gamma)\approx \frac{12\sigma_v^2}{f_s t_m^3}.
\end{equation}

\paragraph{Process-limited (\(\mathcal Q t_m\gg \sigma_v^2\)).}
\begin{equation}
\mathrm{Var}(\hat\gamma)=\frac{\mathcal Q}{t_m}.
\end{equation}

\subsection*{Pressure sensitivity}

In the molecular-flow regime,
\begin{equation}
\gamma=\gamma_P P,
\qquad
P=\frac{\gamma}{\gamma_P}.
\end{equation}
Thus
\begin{equation}
\sigma_P=\frac{\sigma_\gamma}{\gamma_P}.
\end{equation}

The fundamental pressure sensitivity is therefore
\begin{equation}
\sigma_P(t_m)\ge
\max\!\left[
\frac{1}{\gamma_P}\sqrt{\frac{12\sigma_v^2}{f_s t_m^3}},
\;
\frac{1}{\gamma_P}\sqrt{\frac{\mathcal Q}{t_m}}
\right].
\end{equation}

In the process-limited regime, thermal torque noise dominates and
\begin{equation}
\sigma_P = \frac{1}{\gamma_P}\sqrt{\frac{\mathcal Q}{t_m}}.
\end{equation}

Since the underlying thermal torque fluctuations are stationary (white), it is convenient to define an equivalent pressure power spectral density,
\begin{equation}
S_P = \frac{2\mathcal Q}{\gamma_P^2},
\end{equation}
and corresponding amplitude spectral density
\begin{equation}
\delta P_{\mathrm{thermal}} \equiv \sqrt{S_P}
= \frac{\sqrt{2\mathcal Q}}{\gamma_P}.
\end{equation}

For the parameters of our system, this yields
\begin{equation}
\delta P_{\mathrm{thermal}} \approx 2.6\times10^{-11}\ \mathrm{Pa}/\sqrt{\mathrm{Hz}}.
\end{equation}

\section*{Sensitivity to various fundamental physics effects}

\subsection*{Axions}

Axions or axionlike particles (ALPs) are ultralight bosons predicted by several theories beyond the standard models and are among the most considered candidates as constituents of dark matter. If dark matter is made of axions, it will behave as a classical field, which will couple to electron spins in a ferromagnet as an effective fluctuating magnetic field, which generates a torque. Assuming that the totality of dark matter is composed of axions with mass $m_a$, the equivalent magnetic field will be quasimonochromatic with characteristic frequency $\omega_a=m_a c^2/\hbar$, coherence time $t_a=Q_a/\omega_a$ where $Q_a\sim 10^6$, fluctuating (on the timescale $t_a$) direction and magnitude, with rms value $B_{a}\sim g_{ae}\cdot 4 \times 10^{-8}\,\mathrm T$, where $g_{ae}$ is the dimensionless axion-electron coupling. 

Moreover, an important advantage of the MLSR system is that the superconducting shield does not cancel the axion-induced pseudomagnetic field acting on electron spins. This is because superconducting shielding relies on the coupling of real magnetic fields to the motion of charged particles (via the Meissner effect), rather than on the coupling to electron spins as in high-permeability ferromagnetic materials commonly used in atomic-spin experiments \cite{jackson2016magnetic}. As a result, the axion pseudomagnetic field penetrates the superconducting shield without being attenuated, preserving the sensitivity of the system to anomalous spin-dependent interactions.

Beside the method discussed in the main text, a rotor can be used to detect axions even in the free spinning regime. Let us consider an optimal situation in which the rotor is spinning on the $x-y$ plane with magnetic dipole moment $\mu(\cos(\Omega t), \sin(\Omega t),0)$ and the axion field oscillates on the horizontal field as $B_a(0,\cos(\omega_a t),0)$. If $\Omega\sim \omega_a$ there will be a slowly varying net torque along $z$:
\begin{equation}\label{eq:axtorque}
    N_z=\frac{1}{2}\mu B_a \cos{(\omega_a-\Omega)t}.
\end{equation}
The torque will be detected by searching for the induced perturbation of the spinning frequency.

If the rotor is spinning down, $\Omega=\Omega_0 e^{-\gamma t}$, and $\Omega_0 = \omega_a$, this torque will remain coherent as long as the accumulated phase difference remains small:
\begin{equation}\label{eq:phase}
    \Delta \phi = \frac{\Omega_0 \gamma}{2 } t^2 <1
\end{equation}
and will dephase afterwards.
This defines a maximum integration time ${t_{\mathrm{int}}\sim(2/\Omega_0 \gamma)^{1/2}}$.
If we require the integration time to be as long as the axion coherence time, we obtain the interesting relation:
\begin{equation}\label{eq:Qcondition}
    Q > \frac{Q_a^2}{4},
\end{equation}
where $Q=\Omega /2 \gamma$
Remarkably, with $\gamma =5 \times 10^{-7}$ Hz our rotor satisfies this condition for $f_a=\omega_a/2\pi>20$ kHz.
This means that, by simply letting the system spin down, one will automatically perform a sweeped search for axions in the whole range $1$ kHz - $2.3$ MHz with an integration time for each frequency longer than the axion coherence time.

By analogy with the case of a trapped levitated magnet, the signal-to-noise ratio for detecting axions, after averaging over all directions, is \cite{Kalia2024}:  
\begin{equation}\label{eq:SNR}
\mathrm{SNR}=\frac{B_{a}^2}{6 S_{B}} \sqrt{ t_\mathrm{int}\cdot\min(t_\mathrm{int},t_\mathrm{coh})},
\end{equation}
where $t_\mathrm{int}$ is the detector integration time.
We assume that within $t_\mathrm{int}$ the magnetic field noise is dominated by thermal torque noise $S_B= S_N/\mu^2$ with $S_N = 4 k_B T I \gamma$. By setting $\mathrm{SNR}=3$ as detection threshold, we find the potential upper limit on $g_{ae}$ shown in Fig. \ref{fig:axions}, where the pink area corresponds to the current parameters, and the white area to enhanced parameters $R=300$ $\mu$m, $T=1$ K, $\gamma=1\times 10^{-8}$ Hz. The experiment would thus set new laboratory-based bounds in the frequency range kHz-MHz.

\subsection*{High-frequency gravitational waves}

Gravitational waves (GWs) have rapidly moved from the first historical direct detection in 2015 \cite{LIGO2016} into an observational phase. GW signals from known astrophysical objects, such as stellar black holes and neutron stars, are believed to extend up to a maximum frequency of a few kHz, a range covered by large scale interferometers like LIGO. However, a plethora of potential sources have been proposed capable of producing GWs at high frequencies from 10 kHz up to GHz and beyond. The list includes mergers of primordial black holes or exotic compact objects with planetary size, axion superradiance around astrophysical spinning black-holes, and cosmological stochastic background. Further details can be found in a dedicated living review \cite{aggarwal2021}. Different detectors have been proposed to search for high frequency GWs, based on mechanical resonators, levitated particles, microwave cavities or optical interferometers. 

Here we consider a detector based on the torque induced by a GW on a rotor. To this end, let us replace the spherical rotor of our experiment with a quadrupolar rotor, consisting of dumbbell made of two equal microspheres of mass $m$ of radius $R$ with center-to-center distance $2 R$. This can be approximated in practice by an elongated, e.g. cylindrical, rotor. In fact, no net torque is expected for a spherical detector. We assume that the dumbbell is spinning on the $xy$ plane with a phase angle $\alpha(t)= \Omega t$, while a gravitational wave with components $h_{ij}$ at angular frequency $\omega$ propagates along $z$. We ask the question, what is the torque exerted by the gravitational wave on the rotor?

In the long-wavelength approximation, the tidal acceleration acting on a material point with coordinates $x_i$ under the effect of the wave is:
\begin{equation}
\ddot{x}_i(t) =  \frac{1}{2}\,\ddot{h}_{ij}(t)\,x_j .   \label{geodesic}
\end{equation}
For a polarized $h_+$ wave propagating along $z$, the only non-zero components will be $h_{xx} = h(t)$ and $h_{yy} = -h(t)$. The coordinates of the centers of the two masses read:
\begin{align}
&\bm{r}_1 = R(\cos (\Omega t),\sin (\Omega t),0) \\
&\bm{r}_2= -R(\cos(\Omega t),\sin (\Omega t),0) 
\end{align}
For each mass, approximated as a point mass (more refined expressions can be evaluated taking into account the continuous mass distribution), we can use Eq. (\ref{geodesic}) to calculate the force $\bm {F_i} = m \ddot{\bm {r_i}}$ and the torque $\bm{N_i}=\bm{r_i}\times \bm{F_i}$ acting on each mass. 
The only non-vanishing component of the torque is the one along $z$, which after some algebra and summing over the two masses, reads:
\begin{equation}
N_z(t) = m R^2 \ddot{h}(t)\sin(2\Omega t).  \label{torqueGW1}
\end{equation}
This expression oscillates over time, reaching a peak value for the rotation angles $\Omega t = (2 k +1) \pi/4$, that is, when the dumbbell is rotated by $\pi/4$ with respect to the axes $x,y$.
The torque will act on the rotor with maximum efficiency when $2 \Omega = \omega$. In this case the wave is synchronous with the rotor, causing a torque coherent over time, similarly to the case of axions previously discussed.

\begin{figure}
	\centering
	\includegraphics[width=1\textwidth]{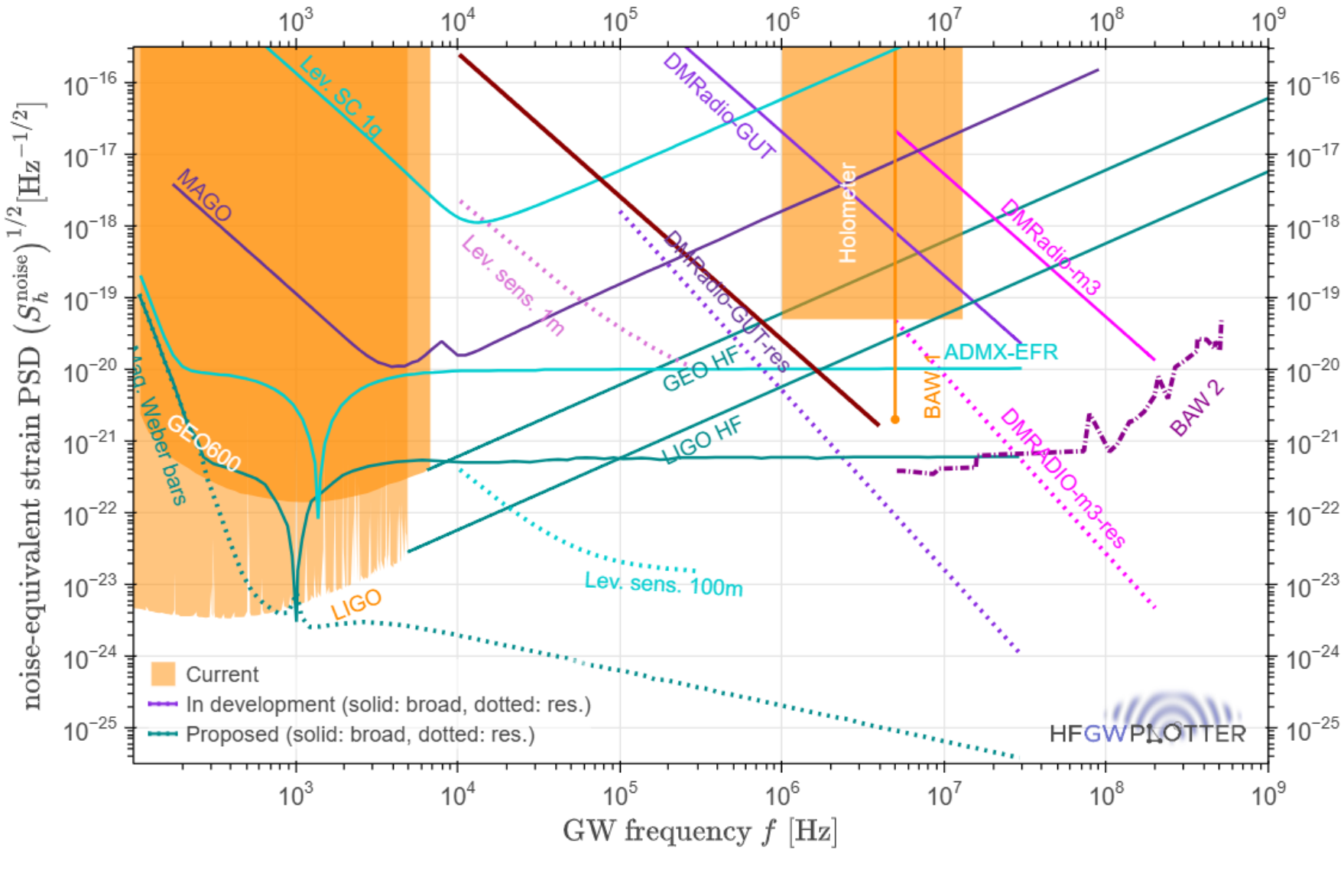}
	\caption{\textbf{Sensitivity to gravitational waves.} Red Curve: gravitational wave strain noise corresponding to the thermal noise of a dumbbell composed of two microspheres as the one demonstrated in the experiment, with a moderate improved of the  experimental parameters $T=1$ K, $\gamma=10^{-8}$ Hz. On the same plot we have shown other relevant actual (orange), under development (purple) or proposed cyan) experiments. Our small and compact detector would be competitive with experiments orders of magnitude bigger in size and cost. Plot produced with the plotter at \it{https://incandenza-01.zdv.uni-mainz.de/Shplot/}}
	\label{fig:GW}
\end{figure}

Under these optimal conditions, the strain noise spectral density is expected to be limited by thermal torque, and can be estimated as:
\begin{equation}
S_h^{1/2} = \frac{S_N^{1/2}}{m R^2 \omega^2}.  \label{torqueGW2}
\end{equation}
where $S_N = 4 k_B T I \gamma$ is the thermal torque noise spectral density. 
Substituting the values obtained in the current experiment (but replacing a sphere with a dumbbell made of two particles of same size), i.e. $R=30 \, \mu$m, $\Omega/2 \pi = 2$ MHz, $\omega=2 \Omega$, $\gamma=5 \times 10^{-7}$ Hz, $T=4.2$ K we estimate $S_{h}^{1/2} = 3 \times 10^{-20} \, \mathrm{Hz}^{-1/2}$. With a plausible improvement to $\gamma = 10^{-8}$ Hz, $T=1$ K we find the red curve in Fig. \ref{fig:GW}. This level of strain noise at MHz frequencies is comparable to that of best bulk acoustic resonators \cite{MAGE2023}, with the significant advantage that our system is much smaller, compact and fully tunable. Therefore, it could be tuned to the frequency of another rotor or bulk acoustic resonator. It would also be competitive with much more massive, demanding, and expensive proposed experiments. An important point that needs to be experimentally addressed is whether the ultralow damping that we have observed in a spherical rotor can be maintained with a dumbbell or cylidrical geometry. As long as the main loss channel is the gas, the effective damping should change only by factors of the order of unity compared to a sphere. However, other potential loss channels will need to be assessed.

\subsection*{Test of spontaneous collapse models}
Spontaneous collapse models have been proposed as an objective universal localization mechanism able to explain the collapse of the wave function of macroscopic objects. This is achieved by introducing stochastic
modifications to the Schr\"odinger equation through the
addition of noise terms that depend on new phenomenological parameters \cite{grw1986}. The presence of such terms results not only in the spontaneous localization of the wavefunction, but also in an unavoidable stochastic force or torque noise acting on any massive mechanical system. Therefore, beside mechanical resonators \cite{vinante2016,carlesso2016}, a highly isolated rotor could also be used to probe or constrain collapse models \cite{carlesso2018}. 

\begin{figure}
	\centering
	\includegraphics[width=0.6\textwidth]{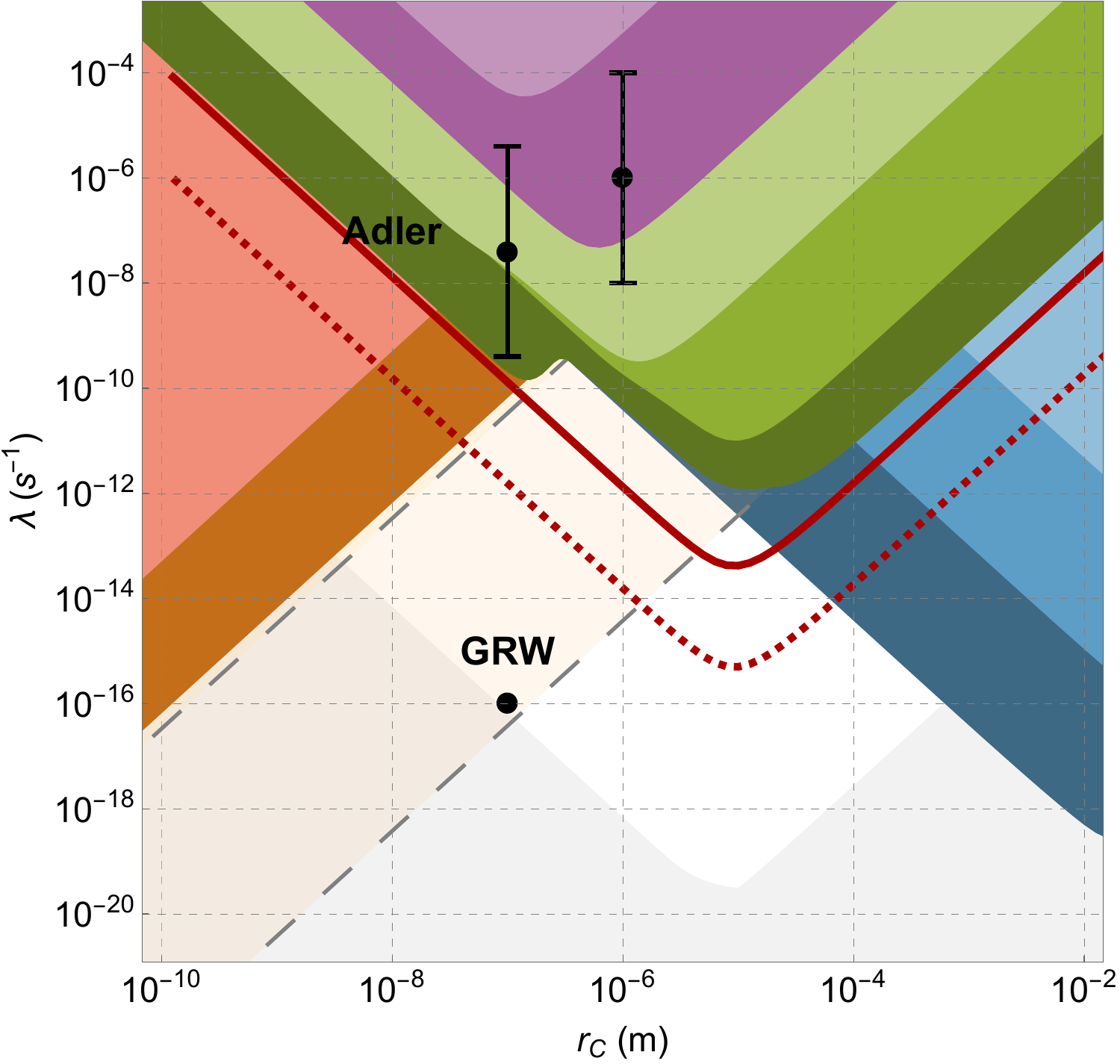}
	\caption{\textbf{Sensitivity to probe the CSL model.} Exclusion plot of the CSL model. The x-axis is the correlation length of the CSL noise $r_C$, the y-axis is the collapse rate for a single nucleon $\lambda$. Blue and green areas on the top part are excluded respectively by gravitational wave detectors \cite{carlesso2016} and ultracold  cantilevers \cite{vinante2016,vinante2020}, light yellow on the left by x-ray emission experiments \cite{arnquist2022}. Typically the limits inferred from x-ray emission are considered less significant because they probe the CSL field at very high frequency. The grey area on the bottom is excluded by theoretical arguments. ``GRW" and ``Adler" refer to theoretically motivated suggestions for the specific values of $r_C$ and $\lambda$. The white area is uncharted. The red and orange thick line show the upper limit on $\lambda$ that would be set by a dumbbell rotor with size similar to our current spherical rotor (two magnets of radius $R=15 \, \mu$m) if the torque noise is measured down to the thermal limit. Solid curve: with parameters already achieved ($T=4.2$ K, $\gamma=5\times10^{-7}$ Hz). Dotted curve: with plausible enhanced parameters ($T=1$ K, $\gamma=1\times10^{-8}$ Hz).}
	\label{fig:CSL}
\end{figure}

Here, we consider the most representative example among collapse models, the so-called Continuous Spontaneous Localization (CSL) model \cite{csl1990}. The model is characterized by two phenomenological parameters that can be constrained by experiments: the correlation length of CSL noise $r_C$, and the collapse rate for a single nucleon $\lambda$. The exclusion plot of the CSL model with existing experiments is shown in Fig. \ref{fig:CSL} \cite{vinante2016,carlesso2016,vinante2017,vinante2020,arnquist2022}.

The CSL-induced rotational diffusion on a perfect spherical rotor is zero \cite{carlesso2018}. Therefore, we consider again a dumbbell rotor as in the previous discussion of GWs, and calculate the CSL-induced torque noise as the incoherent sum of the individual torque noises acting on each sphere, i.e. the radius $R$ times the CSL force noise, using standard formulas \cite{vinante2016}. The upper limit achievable with a rotor with $R=15 \, \mu$m for current parameters ($T=4.2$ K and $\gamma=5\times 10^{-7}$ Hz) is shown as a solid red line, the upper limit with enhanced parameters ($T=1$ K and $\gamma=1\times10^{-8}$ Hz) is shown as a dotted red line. The experiment would thus probe a fully uncharted region of the CSL parameter space.


\end{document}